\documentclass[12pt,draftcls,onecolumn]{IEEEtran}
\usepackage{graphics}
\usepackage{epstopdf}
\usepackage{graphicx}
\usepackage{epsfig,amssymb}
\usepackage{amsmath}
\usepackage{mathrsfs}
\usepackage{color}
\usepackage{array}
\usepackage{amsfonts,booktabs}
\usepackage{lipsum,amsmath,multicol}
\usepackage{graphicx}
\usepackage{subfigure}
\usepackage{times}
\usepackage{pdfsync}
\usepackage{pgfplots}
\usepackage{pgfplotstable}
\usepackage[subnum]{cases}
\usepackage{multirow}
\usepackage{algorithm}
\usepackage{algpseudocode}
\usepackage{float}
\usepackage{setspace}
\usepackage{bm}
\newtheorem{theorem}{Theorem}
\newtheorem{lemma}{Lemma}

\newtheorem{remark}{Remark}

\newcommand{\beq}{\begin{equation}}
\newcommand{\eeq}{\end{equation}}
\newcommand{\beqa}{\begin{eqnarray}}
\newcommand{\eeqa}{\end{eqnarray}}

\newcommand{\paren}[1]{\left(#1\right)}
\newcommand{\sqparen}[1]{\left[#1\right]}

\newcommand{\field}[1]{\ensuremath{\mathbb{#1}}}
\newcommand{\abs}[1]{\left|#1\right|} 
\newcommand{\R}{\ensuremath{\field{R}}} 
\newcommand{\PRP}[1]{\ensuremath{\mathsf{Pr}\left(#1\right)}} 
\newcommand{\ES}[1]{\ensuremath{\mathsf{E}\left[#1 \right]}} 
\newcommand{\e}[1]{\ensuremath{{\rm e}^{#1}}} 

\renewcommand{\vec}[1]{\ensuremath{\boldsymbol{#1}}}

\newcommand{\CES}[2]{\ensuremath{\mathsf{E}\left[\left. #1\right| #2\right]}} 
\newcommand{\CP}[2]{\ensuremath{\mathsf{Pr}\paren{\left.#1\right|#2}}}
\newcommand{\CD}[3]{\ensuremath{p_{#1}\paren{#2\left|#3\right.}}}
\newcommand{\CCDF}[3]{\ensuremath{F_{#1}\paren{#2\left|#3\right.}}}
\newcommand{\INVCCDF}[3]{\ensuremath{F_{#1}^{-1}\paren{#2\left|#3\right.}}}
\newcommand{\CPDF}[3]{\ensuremath{p_{#1}\paren{#2\left|#3\right.}}}
\newcommand{\DF}[2]{\ensuremath{p_{#1}\paren{#2}}}
\newcommand{\DSE}[1]{\ensuremath{\mathsf{H}\sqparen{#1}}}
\newcommand{\CDSE}[2]{\ensuremath{\mathsf{H}\sqparen{#1\left|#2\right.}}}
\newcommand{\MI}[2]{\ensuremath{\mathsf{I}\sqparen{#1;#2}}}
\newcommand{\NLM}[4]{\ensuremath{\Phi_{#1}\paren{#2,#3,#4}}} 
\begin{document}

	\title{A Model Randomization Approach to\\ Statistical Parameter Privacy}
	\author{Ehsan Nekouei, Henrik Sandberg, Mikael Skoglund, and Karl H. Johansson
		\thanks{
			E. Nekouei is with the Department of Electrical Engineering, City University of Hong Kong. E-mail: {\tt enekouei@cityu.edu.hk}. H. Sandberg, M. Skoglund, and K. H. Johansson are with the School of Electrical Engineering and Computer Science, KTH Royal Institute of Technology. E-mail: {\tt \{hsan,skoglund,kallej\}@kth.se}.}}
	\maketitle
	\thispagestyle{empty}
	
	\begin{abstract}
		In this paper, we study a privacy filter design problem for a sequence of sensor measurements whose joint probability density function (p.d.f.) depends on a private parameter. To ensure parameter privacy, we propose a filter design framework which consists of two components: a randomizer and a nonlinear transformation. The randomizer takes the private parameter as input and randomly generates a pseudo parameter. The nonlinear mapping transforms the measurements such that the joint p.d.f. of the filter's output depends on the pseudo parameter rather than the private parameter. It also ensures that the joint p.d.f. of the filter's output belongs to the same family of distributions as that of the measurements. \textcolor{black}{The nonlinear transformation has a feedforward-feedback structure that allows real-time and causal generation of the disguised measurements with low complexity using a recursive structure.} The design of the randomizer is formulated as an optimization problem subject to a privacy constraint, in terms of mutual information, and it is shown that the optimal randomizer is the solution of a convex optimization problem. Using information-theoretic inequalities, we show that the performance of any estimator of the private parameter, based on the output of the privacy filter, is limited by the privacy constraint. The structure of the nonlinear transformation is studied in the special cases of  independent and identically distributed, Markovian, and Gauss-Markov measurements. Our results show that the privacy filter in the Gauss-Markov case can be implemented as two one-step ahead Kalman predictors and a set of minimum mean square error predictors. \textcolor{black}{The Kalman predictors significantly reduce the complexity of computing the disguised measurements. A numerical example on occupancy privacy in a building automation system illustrates the approach.}
	\end{abstract}

	\section{Introduction}
	\subsection{Motivation}
	Networked control systems are omnipresent in our daily lives by providing essential services such as intelligent transportation, smart grid, and intelligent buildings. Sensors are crucial components of any   system as they provide critical information that can be used for control, diagnosis and monitoring purposes. However, sensor measurements in a networked system typically contain information about private variables. Thus, directly revealing the measurements to untrusted parties may expose the system to the risk of privacy loss.  For example, the occupancy level of a building, which is a highly private variable, can be inferred from the $\rm CO_2$ and temperature measurements of the building \cite{EBVWJ15}. Privacy breaches may have  negative consequences such as reputation damage and financial losses due to lawsuits. The sheer importance of privacy has motivated numerous research efforts to develop privacy-preserving solutions for networked control systems. 
	\subsection{Related Work}
	A variety of problems related to privacy and networked estimation and control have been studied in the literature. The privacy aspect of the hypothesis testing problem as well as various solutions for privacy-aware hypothesis testing have been studied in the literature,  \emph{e.g.,}  \cite{HTS16, ST16, HT17, LSTC16}. The authors in \cite{LO15} considered a multi-sensor hypothesis testing problem wherein a fusion center receives the decisions of a set of sensors and an adversary overhears the local decisions of a subset of sensors. They studied the optimal privacy-aware hypothesis testing rule that minimizes the Bayes risk subject to a privacy constraint at the adversary.  In \cite{LO17}, the authors considered a similar set-up  and  investigated the optimal privacy-aware design of the Neyman-Pearson test. 
	
 The authors in \cite{MO18} studied the optimal privacy filter design problem for a public Markov chain that is correlated with a private Markov chain. Tanaka \emph{et al. } in \cite{TSSJ17} considered a linear Gaussian plant in which the sensor measurements are communicated with an untrusted cloud controller. They studied the optimal privacy filter design problem subject to a constraint on the privacy of the states of the plant. The authors of \cite{V13} studied the optimal privacy-aware control law for a Markov decision process subject to a privacy constraint on an adversary which has access to the input and output of the Markov chain. 	We note that the information-theoretic approach to privacy has been extensively studied in the literature, \emph{e.g.,} \cite{KSK17}, \cite{BWI16}, \cite{CF12}, \cite{MS15}. In such an approach, one observes a public random variable which is correlated with a private variable. Here, the objective is to generate a degraded version of the public variable such that the distortion due to the privacy filter is minimized subject to an information-theoretic constraint on the privacy. The interested reader is referred to \cite{NTSJ19} for an overview of information-theoretic approaches to privacy in estimation and control.
	
Privacy-aware solutions to estimation, filtering, and average consensus problems have been developed in the literature based on the notion of differential privacy. The authors in \cite{NP14} developed a framework for privacy-aware filtering of the measurements of a dynamical system using the concept of differential privacy. Sandberg \emph{et al.}  \cite{SDT15} studied the state estimation in an electricity distribution network subject to a constraint on the consumers' privacy. Privacy-aware average consensus algorithms were developed in \cite{NTC17} and \cite{MM17} to guarantee the privacy of agents' initial states. The authors of  \cite{WHMD17} developed a framework based on differential privacy to address the privacy of the initial state as well as the way-points of each agent in a distributed control problem. 
	
The statistical parameter privacy problem has been studied in \cite{BSP18} and \cite{ZS20}. Bassi \emph{et al.} in \cite{BSP18} studied the statistical parameter privacy of an independent and identically distributed (i.i.d.) sequence of random variables where their common p.d.f. depends on a private parameter. In their set-up, the privacy filter consists of a randomly selected stochastic kernel which generates a random output based on each observed random variable. They characterized the leakage level of private information under the Bayes statistical risk as the privacy measure. The authors in \cite{ZS20} studied the optimal design of controller and privacy filter for a linear Gaussian plant. The system dynamics should be kept private from an adversary interested in inferring the system dynamics based on the state measurements and control inputs. Assuming Fisher information as privacy metric, they showed that the optimal privacy filter is in the form of a state-dependent Gaussian stochastic kernel.

	\subsection{Contributions}
	In this paper, we consider a sequence of measurements, observed by a sensor, whose joint p.d.f. depends on a private parameter. To ensure   parameter privacy of the measurements, we propose a  filter design framework which consists of two parts: a randomizer and a nonlinear transformation.  The randomizer takes the private parameter as input and randomly generates a pseudo parameter. The nonlinear transformation alters the measurements such that the joint p.d.f. of the filter's output is characterized by the pseudo parameter rather than the private parameter. The nonlinear transformation also ensures that the joint p.d.f. of the filter's output belongs to the same family of  distributions as the measurements. \textcolor{black}{The nonlinear transformation has a feedforward-feedback structure which enables real-time and causal computation of the disguised measurements with low complexity. }

	In our set-up, the randomizer is designed by minimizing the average distortion due to the privacy filter subject to a privacy constraint in terms of the mutual information between the private and the pseudo parameters.  Using information-theoretic inequalities, we show that the performance of any estimator of the private parameter based on the output of the privacy filter is limited by the privacy constraint. 
		We investigate the structure of the nonlinear transformation for independent and identically distributed (i.i.d.), Markovian and Gauss-Markov measurements. Our results show that the structure of the nonlinear transformation involves two one-step-ahead Kalman predictors in the Gauss-Markov case. \textcolor{black}{The Kalman predictors significantly reduce the complexity of generating the disguised measurements.} The result for i.i.d. measurements has appeared in \cite{BNSJ20}.
		
		  Different from \cite{BNSJ20} and \cite{BSP18}, the current paper develops a privacy filter design framework without imposing the i.i.d. assumption on the joint p.d.f. of the measurements.  In our set-up, the sensor measurements are processed by a nonlinear transformation, rather than a stochastic kernel, which ensures that the distribution of the output of privacy filter belongs to the same family of distributions as the measurements. This requirement is not guaranteed by the frameworks in \cite{BSP18} and \cite{ZS20} since stochastic kernels significantly alter the distribution of the measurements.
	 
	   Our mutual information privacy metric is fundamentally different from the Fisher information metric in \cite{ZS20}, in the sense that the mutual information provides a lower bound on the error probability of any arbitrary estimator of private variables whereas Fisher information provides a lower bound on the mean square error of any unbiased estimator of private parameters.
	  \subsection{Outline}
	This paper is organized as follows. Section \ref{Sec: SM} describes our system model and standing assumptions. Section \ref{Sec: MR} introduces the model randomization approach to parameter privacy.  Section \ref{Sec: SC} investigates the structure of the nonlinear transformation in special cases. Section \ref{Sec: NR} presents the numerical results followed by the concluding remarks in Section \ref{Sec: Conc}. 
	
	\section{Problem Formulation}\label{Sec: SM}
	\subsection{System Model and Objectives}
Consider a sensor that measures the stochastic process $Y_k$ over the time-horizon $k=1,\dots,T$ where $Y_k=\left[Y_k^1,\dots ,Y_k^d\right]^\top$ is a $d$-dimensional random vector.  We assume that the joint p.d.f. of the measurements over the time-horizon $1,\dots,T$ is parameterized by  $\Theta$ which takes values in $\vec{\Theta}=\left\{\theta_1,\dots,\theta_m\right\}$ with probability mass function $\PRP{\Theta=\theta_i}=p_i$. The value of the parameter $\Theta$ is fixed during the time-horizon $k=1,\dots,T$. 
The joint p.d.f. of $\left\{Y_k\right\}_{k=1}^T$ is denoted by  $p_{\theta_i}\paren{y_1,\dots,y_T}$ when $\Theta$ is equal to $\theta_i$.   We refer to  $p_{\theta}\paren{y_1,\dots,y_T}$  as the \emph{statistical model} of the measurements which belongs to the family of p.d.f.s
\begin{align}
\mathcal{M}=\left\{p_\theta\paren{y_1,\dots,y_T}\right\}_{\theta}\nonumber.
\end{align}  
			\begin{figure*}[t!]
			\centering
			\includegraphics[scale=0.5]{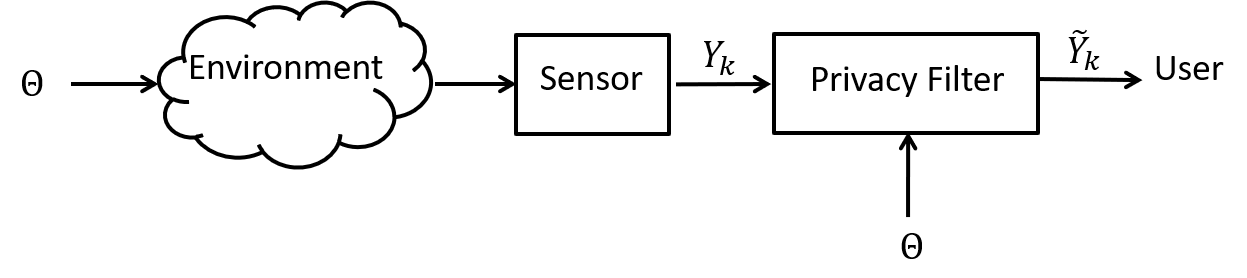}
			\caption{The information sharing with an untrusted user via a privacy filter.}
			\label{Fig: Sensor}
		\end{figure*}	
  	\begin{figure}[t!]
  	\centering
  	\includegraphics[scale=0.65]{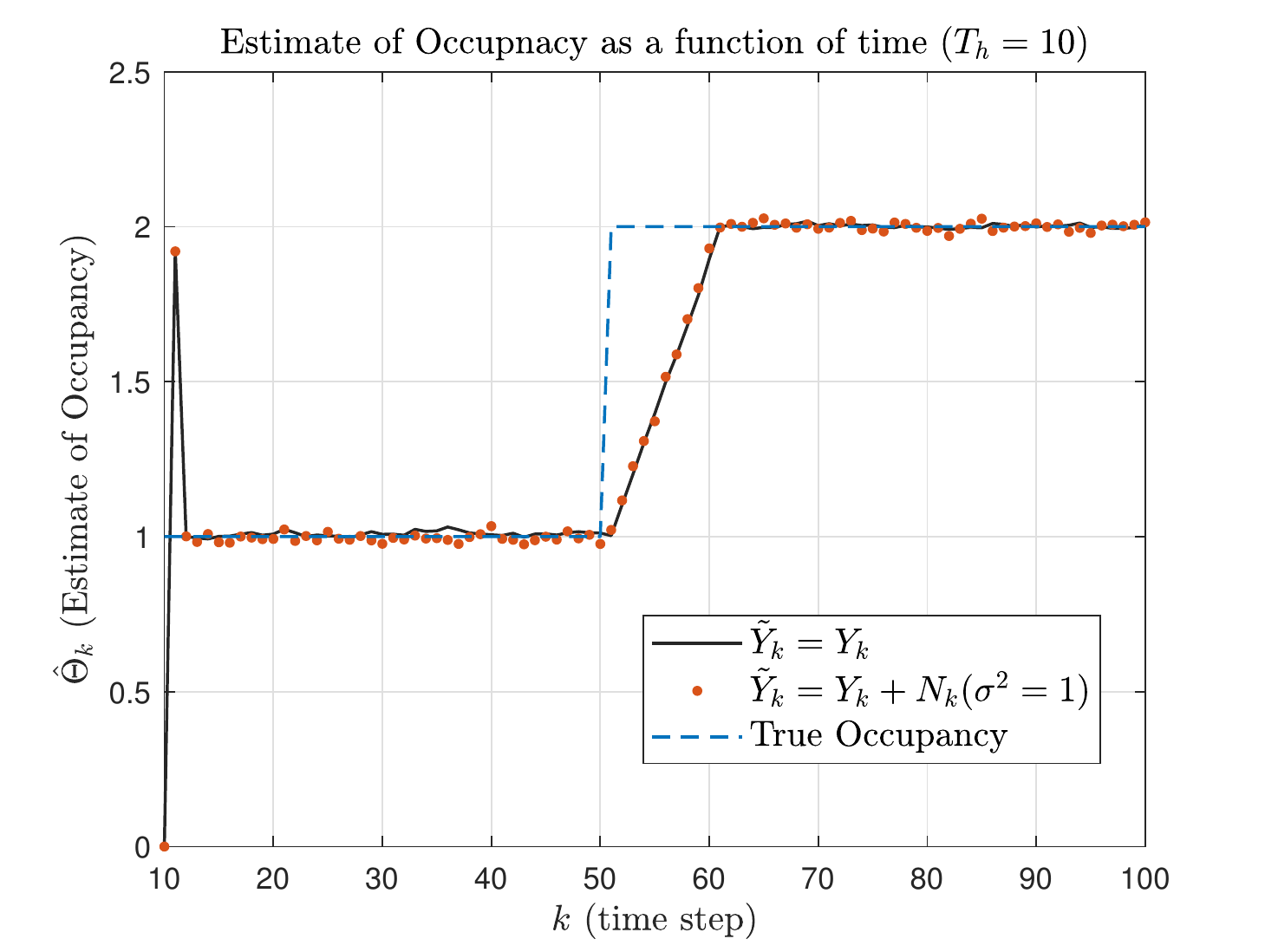}
  	\caption{\textcolor{black}{Estimates of occupancy ($\Theta$) as a function of time for different levels of occupancy}.}
  	\label{fig: F0}
  \end{figure}

   We assume that $\Theta$ carries private information and the sensor sequentially communicates its measurements with an untrusted party, hereafter named the ``user", for monitoring, control or storage purposes. Directly revealing the measurements to the user will result in the loss of private information since the true value of $\Theta$ can be inferred from the measurements. A common approach for ensuring privacy is to use a privacy filter that mediates the information sharing between the sensor and user as shown in Fig. \ref{Fig: Sensor}. In this paper, we develop a privacy filter design framework that achieves the following three objectives:
   \begin{enumerate}
   	\item It ensures that the output of the privacy filter accurately represents the sensor measurements. 
   	\item It guarantees that an adversary with access to the output of the privacy filter cannot reliably infer the value of $\Theta$. 
   	\item The joint p.d.f. of the output of the privacy filter belongs to the family of distributions  $\mathcal{M}$.
   \end{enumerate}

 	In general, the output distribution of a privacy filter might be arbitrarily complex. Objective 3 avoids this situation by ensuring that the joint p.d.f. of the privacy filter's output belongs to the same family of distributions as that of the measurements. This is especially important when the filter's output is used to perform computations, such as filtering, whose complexity depends on the underlying distribution of the data.
 	       
         Note that adding noise to the measurements may not ensure the privacy of the parameter $\Theta$. To highlight this point, assume that the $Y_k$'s are i.i.d. according to a Gaussian distribution with mean $\Theta$ and unit variance. Let $\tilde{Y}_k=Y_k+N_k$ denote the shared information with the user where $\left\{N_k\right\}_k$ is a sequence of zero mean  independent and identically distributed random variables. Using the law of large numbers, we have 
        \begin{align}
        \frac{1}{T}\sum_{k=1}^T\tilde{Y}_k\rightarrow\Theta\nonumber
        \end{align}
        almost surely as $T$ tends to infinity, which indicates that the user can reliably estimate the private parameter when $T$ is large. Hence, the noise addition mechanism does not ensure the privacy of $\Theta$. In this example, the statistical model of the measurements belongs to the family of distributions 
        \begin{align}
       \mathcal{M}=\left\{\frac{1}{\sqrt{2\pi}}\e{\frac{1}{2}\paren{x-\theta}^2}\right\}_{\theta\in\R}\nonumber
        \end{align} 
         where $\theta$ takes values in the set of real numbers. If the additive noise terms $N_k$s are Laplacian distributed \cite{DR14}, objective 3 is not satisfied and the p.d.f. of the shared information with the user becomes intractable.

\begin{remark}
	Although, we assume that $\Theta$ is fixed, our results can be easily extended to the case that $\Theta$ is time-varying, \emph{e.g.,} when the private parameter is given by $\Theta=\left[\Theta_1,\dots,\Theta_T\right]$ where $\Theta_k$ is the parameter that characterizes the conditional distribution of $Y_k$ given $Y_1,\dots,Y_{k-1}$.
\end{remark}

	\subsection{Motivating Example: Building Automation}\label{Sec: ME}
	Consider a building automation application in which a sensor measures the ${\rm CO}_2$ level of a room over the time horizon $k=1,\dots,T$. The response of $\rm CO_2$ concentration to the presence of humans can be modeled as 	\begin{align}
	{X}_{k+1}&=a{X}_k+{W}_k+b\Theta\nonumber\\
	{Y}_{k}&={X}_k+{V}_k\nonumber
	\end{align}
	where $a\in\left(0,1\right)$ and $b$ are  constants, ${X}_k$ denotes the ${\rm CO}_2$ level at time-step $k$, $Y_k$ represents the sensor measurement,  ${W}_k$ denotes the external disturbance, $V_k$ denotes the measurement noise and $\Theta\in\left\{0,\dots,L\right\}$ denotes the occupancy level of the room, \emph{i.e.,} $\Theta=i$ indicates that there are $i$ persons in the room during the horizon $k=1,\dots,T$. 
	
	In building automation, it is common to transmit the sensor measurements over communication networks for control or monitoring purposes. Note that $\Theta$ is the private statistical parameter of the shared information $\left\{Y_k\right\}_{k=1}^T$. When the $\rm CO_2$ measurements are accessible by untrusted parties, \emph{e.g.,} a hacker or a ``cloud", the occupancy level, which carries private information, can be inferred from the measurements. 
	
We examine the privacy of the occupancy information in two cases: ($i$) when $\rm CO_2$ measurements are directly shared with the user, ($ii$) when a noise addition mechanism is employed to ensure occupancy privacy. To this end, we first construct an estimator of the occupancy level based on the shared information with the user. Let $\tilde{Y}_k$ denote the shared information at time-step $k$ and define the averages $\bar{Y}_k$ and $\bar{\bar{Y}}_k$ as 
	\begin{align}
	\bar{Y}_k&=\frac{1}{T_h}\sum_{i=k-\left(T_h-1\right)}^k\tilde{Y}_i\\
	\bar{\bar{Y}}_k&=\frac{1}{T_h}\sum_{i=k-T_h}^{k-1}\tilde{Y}_i\nonumber
	\end{align}
	for $k\geq T_h+1$ where $T_h$ is the window size. Let $\hat{\Theta}_k$ denote the estimator of occupancy based on the shared information up to time-step $k$, $\tilde{Y}_1,\dots,\tilde{Y}_k$, which is defined as 
	\begin{align}\label{Eq: O-E}
	\hat{\Theta}_k=\frac{\bar{Y}_k-\bar{\bar{Y}}_k}{a}.
	\end{align}

	Fig. \ref{fig: F0} shows the output of the occupancy estimator, as a function of time, when the true occupancy level is equal to 1 or 2. The solid lines in Fig. \ref{fig: F0} show the occupancy estimates when the estimator has access to the $\rm CO_2$ measurements, \emph{i.e.,} $\tilde{Y}_k=Y_k$. The dashed lines in this figure represent the occupancy estimates under a noise addition privacy mechanism wherein Gaussian noise is added to the measurements. In this case, the estimator has access to $\tilde{Y}_k=Y_k+N_k$ where $\left\{N_k\right\}_k$ is a sequence of independent and identically distributed (i.i.d.) Gaussian random variables with zero mean and unit variance. In our simulations, $\{W_k,V_k\}_k$ is assumed to be  a sequence of i.i.d. Gaussian random variables with zero mean and variance $0.1$.  
	
	According to Fig. \ref{fig: F0}, the occupancy estimator in \eqref{Eq: O-E} can reliably infer the occupancy  level even if the noise addition privacy mechanism is employed. This observation confirms that directly sharing sensor measurements with an entrusted party  might result in the loss of private information. Moreover, based on Fig. \ref{fig: F0}, noise addition mechanisms may not be capable of ensuring statistical parameter privacy. 
\subsection{Notations and Assumptions}
The shorthand notation $Y_{1:k}$ is used to represent the sequence of random variables $Y_1,\dots,Y_k$. The realization of $Y_{1:k}$ is denoted by $y_{1:k}$. The shorthand notation $Y^{1:l}_k$ denotes the collection of the first $l$ components of $Y_k$. The $l$th component of $Y_k$ is denoted by $Y^l_k$. When $\Theta$ is equal to $\theta$, the joint p.d.f. of $Y_{1:k}$ is denoted by $p_\theta\paren{y_1,\dots,y_k}$ which is assumed to be non-zero almost everywhere in $\R^{k\times d}$. 

The conditional p.d.f. of $Y^l_k$ given $\left\{Y^{1:l-1}_k=y^{1:l-1}_k,Y_{1:k-1}=y_{1:k-1},\Theta=\theta_i\right\}$ is represented by 
\begin{align}
p_{\theta_i}\paren{x\left|y^{1:l-1}_k,y_{1:k-1}\right.},\nonumber
\end{align}
 where its corresponding cumulative distribution function (c.d.f.)  is assumed to be absolutely continuous with respect to the Lebesgue measure. The conditional c.d.f. of $Y^l_k$ given \newline \textcolor{black}{$\left\{Y^{1:l-1}_k=y^{1:l-1}_k,Y_{1:k-1}=y_{1:k-1},\Theta=\theta_i\right\}$} is defined as 
\begin{align}\label{Eq: CDF1}
\CCDF{l,k,\theta_i}{z}{y^{1:l-1}_k,y_{1:k-1}}\!\!=\!\!\!\int_{-\infty}^{z}\!\!\!\!\!\!p_{\theta_i}\paren{x\left|y^{1:l-1}_k,y_{1:k-1}\right.}dx.
\end{align} 
 The inverse function of $\CCDF{{l,k,\theta_i}}{\cdot}{{y}^{1:l-1}_k,{y}_{1:k-1}}$ is denoted by $\INVCCDF{{l,k,\theta_i}}{\cdot}{{y}^{1:l-1}_k,{y}_{1:k-1}}$. Note that, for all $\theta_i$, the transformation $\CCDF{l,k,\theta_i}{\cdot}{\cdot}$ is a mapping from $\R^{l+(k-1)d}$ to $\left[0,1\right]$ which is increasing in its first argument when its second argument is fixed.
 
  The following conventions are adopted in the rest of this paper:
  \begin{align}
    \CCDF{1,1,\theta_i}{z}{y^{1:0}_k,y_{1:0}}&=F_{1,1,\theta_i}\paren{z},\nonumber\\
  \CCDF{l,1,\theta_i}{z}{y^{1:l-1}_1,y_{1:0}}&=\CCDF{l,1,\theta_i}{z}{y^{1:l-1}_1},\nonumber\\
    \CCDF{1,k,\theta_i}{z}{y^{1:0}_k,y_{1:k-1}}&=\CCDF{1,k,\theta_i}{z}{y_{1:k-1}},\nonumber
  \end{align}
  where $F_{1,1,\theta_i}\paren{z}$ is the conditional c.d.f. of $Y^1_1$ given $\Theta=\theta_i$, $\CCDF{l,1,\theta_i}{z}{y^{1:l-1}_1}$ is the conditional c.d.f. of  $Y^l_1$ given $\left\{Y^{1:l-1}_1=y^{1:l-1}_1,\Theta=\theta_i\right\}$ and  $\CCDF{l,k,\theta_i}{z}{y_{1:k-1}}$ is the conditional c.d.f. of  $Y^1_k$ given \textcolor{black}{$\left\{Y_{1:k-1}=y_{1:k-1},\Theta=\theta_i\right\}$}. 
	\section{The Model Randomization Approach}\label{Sec: MR}
\begin{figure}[t!]
		\centering
		\includegraphics[scale=0.5]{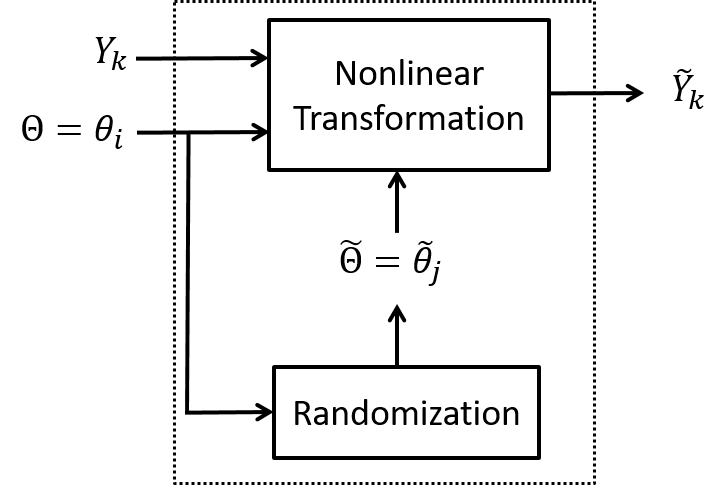}
		\caption{The structure of the proposed privacy filter.}
		\label{fig: F1}
	\end{figure}

	In this section, we discuss the model randomization framework for ensuring the statistical parameter privacy of the sensor measurements. In this framework, the privacy filter consists of two components: a randomizer and a nonlinear transformation, as shown in Fig. \ref{fig: F1}, which collectively attain the objectives 1, 2, and 3 in Section \ref{Sec: SM}. The randomizer takes the value of the private variable $\Theta$ as input and generates a realization of the  \emph{pseudo parameter} $\tilde{\Theta}$ which remains constant during the horizon. The pseudo parameter takes values in the set $\vec{\tilde{\Theta}}=\left\{\tilde{\theta}_1,\dots,\tilde{\theta}_{\tilde{m}}\right\}$.  At each time-step $k$, the nonlinear transformation generates  $\tilde{Y}_k$ based on the sensor measurement at time-step $k$ and the values of $\Theta$ and $\tilde{\Theta}$. Then, $\tilde{Y}_k$  is revealed to the user. 
	
	In this section, the design of the privacy filter is studied without imposing any special structure on the joint p.d.f. of the sensor measurements.  We start by discussing the structure of the nonlinear transformation in the next subsection. Then, the optimal design of the randomizer is discussed, followed by the privacy analysis of $\Theta$ under the proposed framework.

	\subsection{Nonlinear Transformation}
	
\begin{figure*}
		\centering{\includegraphics[scale=0.5]{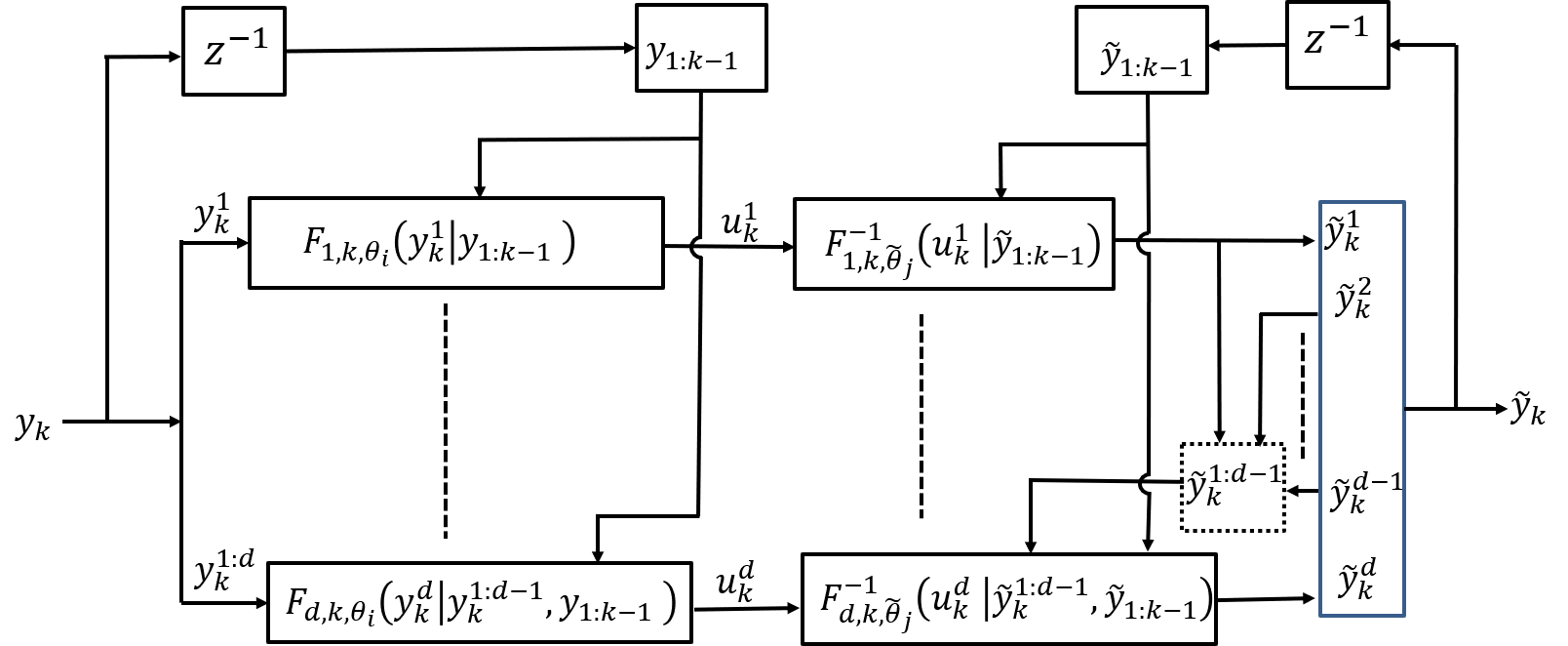}}
		\caption{The structure of the nonlinear transformation at time-step $k$.}\label{Fig: GS}
	\end{figure*}

	 Let $\tilde{y}_k=\left[\tilde{y}_k^1,\dots,\tilde{y}_k^d\right]^\top$ denote the realization of the output of the privacy filter at time $k$. Assuming $\Theta=\theta_i$ and  $\tilde{\Theta}=\tilde{\theta}_j$, the nonlinear transformation at time-step $k$ generates $\tilde{y}^l_k$, \emph{i.e.,} the $l$th entry of $\tilde{y}_k$, according to
\begin{align}\label{Eq: NLT-1}
\tilde{y}^l_k&=\INVCCDF{{l,k,\tilde{\theta}_j}}{u^l_k}{\tilde{y}^{1:l-1}_k,\tilde{y}_{1:k-1}},
\end{align}
where $u^l_k$ is given by
\begin{align}\label{Eq: NLT-2}
u^l_k&=\CCDF{l,k,\theta_i}{y^l_k}{{y}_k^{1:l-1},y_{1:k-1}},
\end{align}
	and $\CCDF{l,k,\theta_i}{\cdot}{{y}_k^{1:l-1},y_{1:k-1}}$ is the c.d.f. of $Y^l_k$ given \textcolor{black}{$\left\{Y^{1:l-1}_k=y^{1:l-1}_k,Y_{1:k-1}=y_{1:k-1},\Theta=\theta_i\right\}$}  defined in \eqref{Eq: CDF1}. According to \eqref{Eq: NLT-1} and \eqref{Eq: NLT-2}, the output of the privacy filter at time-step $k$ is generated based on the history of the measurements up to time-step $k$. Moreover, the $l$th entry of $\tilde{y}_k$ is generated using its first $l-1$ entries, \emph{i.e.,} $\tilde{y}^{1:l-1}_k$, which implies that the entries of the filter's output are generated sequentially. The structure of the nonlinear transformation at time-step $k$ is illustrated in Fig. \ref{Fig: GS}.
 
\textcolor{black}{The \emph{feedforward-feedback} structure of the nonlinear transformation, in Fig. \ref{Fig: GS}, is a unique aspect of the proposed privacy filter. The feedforward component computes $u^i_k$'s whereas the feedback component computes $\tilde{y}^i_k$'s based on the past outputs of the filter. This structure allows us to cast the proposed privacy filter as a {\em dynamical system} enabling recursive computation of the p.d.f.s $p_{\theta_i}\paren{y_k\left|y_{1:k-1}\right.}$ and $p_{\theta_j}\paren{\tilde{y}_k\left|\tilde{y}_{1:k-1}\right.}$. Under the proposed scheme, $p_{\theta_i}\paren{y_k\left|y_{1:k-1}\right.}$ and $p_{\theta_j}\paren{\tilde{y}_k\left|\tilde{y}_{1:k-1}\right.}$ can be computed recursively over time to reduce the computational cost of generating the disguised measurements. The recursive structure significantly reduces the computational cost when the measurements are generated by a linear Gaussian system (see Subsection \ref{Subsec: GM} for more details).}
	
\textcolor{black}{In certain applications, it is important to generate the disguised measurements in real-time, \emph{i.e.,} $\tilde{y}_k$  must be causally generated using the sensor measurements up to time $k$ rather than all the sensor measurements $Y_1,\dots,Y_T$. The recursive structure of the proposed privacy filter enables the causal (real-time) generation of the disguised measurements. }

	The next theorem studies the joint probability density function (p.d.f.) of the output of the nonlinear transformation. 
	\begin{theorem}\label{Theo: Dist_Gen}
		Consider the nonlinear transformation above and assume that the joint p.d.f. of the measurements belongs to the family of distributions $\mathcal{M}=\left\{p_\theta\paren{y_1,\dots,y_T}\right\}_{\theta}$. Given $\Theta=\theta_i$ and $\tilde{\Theta}=\tilde{\theta}_j$, the joint p.d.f. of the filter's output is $p_{\tilde{\theta}_j}\paren{\tilde{y}_1,\dots,\tilde{y}_T}$ for all $i,j$. 
	\end{theorem}
	\begin{IEEEproof}
		See Appendix \ref{App: Dist_Gen}.
	\end{IEEEproof}      
	According to Theorem \ref{Theo: Dist_Gen}, the joint p.d.f. of the output of the privacy filter over the horizon $1,\dots,T$ is characterized by the value of the pseudo parameter $\tilde{\Theta}$ rather than the value of the private parameter $\Theta$. Moreover, the nonlinear transformation ensures that the joint p.d.f. of the filter's output also belongs to the family of probability distributions $\mathcal{M}$. Under the proposed framework, the statistical model of the filter's output, \emph{i.e.,} $p_{\tilde{\theta}}\paren{\tilde{y}_1,\dots,\tilde{y}_T}$, is randomly chosen from the set of probability distributions $\left\{p_{\tilde{\theta}}\paren{\tilde{y}_1,\dots,\tilde{y}_T}\right\}_{\tilde{\theta}\in\vec{\tilde{\Theta}}}$ where $\vec{\tilde{\Theta}}=\left\{\tilde{\theta}_1,\dots,\tilde{\theta}_{\tilde{m}}\right\}$ is the support set of the randomizer's output. Hence, we refer to this framework as the model randomization approach to the parameter privacy.
	
	\subsection{Randomizer}\label{Subsec: Randmizer}
	The parameter $\tilde{\Theta}$ is selected from the set $\vec{\tilde{\Theta}}=\left\{\tilde{\theta}_1,\dots,\tilde{\theta}_{\tilde{m}}\right\}$ using a randomized mapping. More precisely, given $\Theta=\theta_i$, the value of $\tilde{\Theta}$ is randomly generated according to 
	\begin{eqnarray}
	\tilde{\Theta}=
	\left\{
	\begin{array}{ccc}
	\tilde{\theta}_1   &\text{w.p.} \quad P_{1i}, & \nonumber\\  
	\vdots&\vdots&\text{if} \quad \Theta=\theta_i\nonumber\\
	\tilde{\theta}_{\tilde{m}}   & \text{w.p.}\quad  P_{\tilde{m}i},& \nonumber\\
	\end{array}
	\right.\nonumber
	\end{eqnarray}
	where $\sum_jP_{ji}=1$ for all $i$. Thus, the randomizer with probability $P_{ji}$ selects $\tilde{\theta}_j$ as the value of $\tilde{\Theta}$ when $\Theta$ is equal to $\theta_i$. The set of randomization probabilities $\left\{P_{ji}\right\}_{ji}$ are designed such that the accuracy of the output of the privacy filter is maximized and simultaneously a desired privacy level is achieved.  
	
	Due to the nonlinear transformation, the output of the privacy filter might be different from its input. To quantify the difference between the input and the output of the filter, we define the average distortion between $Y_{1:T}$ and $\tilde{Y}_{1:T}$ as  
	\begin{align}
	\frac{1}{T}\sum_{k=1}^T\ES{d\paren{Y_k,\tilde{Y}_k}}\nonumber
	\end{align}
	where the distortion function  $d\paren{\cdot,\cdot}:\R^d\times\R^d\rightarrow \R_+$ captures the deviation of the output of the privacy filter from its input.
	
	We consider the mutual information between the private parameter and pseudo parameter as the privacy metric. Let $\MI{\Theta}{\tilde{\Theta}}$ denote the mutual information between $\Theta$ and $\tilde{\Theta}$ which can be written as
	\begin{align}
	&\MI{\Theta}{\tilde{\Theta}}\nonumber\\
	&=\sum_{i,j}\PRP{\Theta=\theta_i,\tilde{\Theta}=\tilde{\theta}_j}\log\frac{\PRP{\Theta=\theta_i,\tilde{\Theta}=\tilde{\theta}_j}}{\PRP{\Theta=\theta_i}\PRP{\tilde{\Theta}=\tilde{\theta}_j}}.\nonumber
	\end{align}
  Note that $\MI{\Theta}{\tilde{\Theta}}$  captures the amount of information that can be inferred about the private parameter by observing the pseudo parameter. When the mutual information between $\Theta$ and $\tilde{\Theta}$ is zero, the pseudo parameter has no information about the private parameter. Also, the mutual information between $\Theta$ and $\tilde{\Theta}$ achieves its maximum level when $\Theta$ can be correctly inferred  by observing $\tilde{\Theta}$. Thus, a relatively large level of $\MI{\Theta}{\tilde{\Theta}}$ indicates that ${\Theta}$ can be reliably inferred from $\tilde{\Theta}$. 
	
The optimal randomization probabilities are obtained by minimizing the average distortion subject to a privacy constraint. More precisely, the optimal randomization probabilities are the solution of the following optimization problem
	\begin{align}\label{Eq: OP}
	\underset{ \left\{P_{ji}\right\}_{j,i} }{\rm minimize} \quad&    \frac{1}{T}\sum_{k=1}^T\ES{d\paren{Y_k,\tilde{Y}_k}}\nonumber\\
	& P_{ji}\geq 0, \forall i, j\nonumber\\
	&\sum_jP_{ji}=1, \quad \forall i\nonumber\nonumber\\
	& \MI{\Theta}{\tilde{\Theta}}\leq {I}_0.
	\end{align}
	where the second constraint enforces the law of total probability and the last constraint imposes an upper bound on the mutual information between the private parameter and the pseudo parameter.  We refer to the last constraint as the privacy constraint since it limits the amount of information that can be inferred about ${\Theta}$ based on $\tilde{\Theta}$. In what follows, we refer to $I_0$ as the leakage level of private information. 

Next theorem investigates the structural properties of the optimization problem  \eqref{Eq: OP}.
	\begin{theorem}\label{Theo: Conv}
		The objective function in \eqref{Eq: OP} is linear in the randomization probabilities. Also, the privacy constraint is a convex constraint. 
	\end{theorem}
	\begin{IEEEproof}
		See Appendix \ref{App: Conv}.
	\end{IEEEproof}
	Theorem \ref{Theo: Conv} shows that the optimization problem \eqref{Eq: OP} is a convex optimization problem. Hence, the optimal randomization probabilities can be computed efficiently. 
	\begin{remark}
	\textcolor{black}{In practice, the elements of $\vec{\tilde{\Theta}}$ can be selected using a nested optimization problem where the optimal elements of $\vec{\tilde{\Theta}}$ and the optimal randomization probabilities are computed recursively such that the total distortion is minimized while the privacy constraint is met.
		Although this results in a non-convex optimization problem, one can obtain a locally optimal solution by alternating between the nested optimization problems.}
	\end{remark}

\begin{remark}
	\textcolor{black}{It may not be always possible to find a closed-form expression for the total distortion due to the structure of the distortion function and the distribution of the sensor measurements. However, it is straightforward to approximate the total distortion accurately using  Monte-Carlo simulations.}
\end{remark}
	\subsection{Privacy level of $\Theta$}\label{Subsec: Privacy}
	In this subsection, we study the privacy level of the parameter $\Theta$ under the proposed framework. To this end, let $\hat{\Theta}\paren{\tilde{Y}_{1:T}}$ denote an arbitrary estimator of $\Theta$,  based on the output of the privacy filter over the horizon $1,\dots,T$, which is defined as a mapping from $\R^{T\times d}$ to $\vec{\Theta}=\left\{\theta_1,\dots,\theta_m\right\}$. Next theorem establishes a lower bound on the error probability of the estimator $\hat{\Theta}\paren{\tilde{Y}_{1:T}}$.
	\begin{theorem}\label{Theo: EP-LB}
		Let $\PRP{\Theta\neq \hat{\Theta}\paren{\tilde{Y}_{1:T}}}$ denote the error probability of the estimator $\hat{\Theta}\paren{\tilde{Y}_{1:T}}$. Then, we have 
			\begin{align}
		\PRP{\Theta\neq\hat{\Theta}\paren{\tilde{Y}_{1:T}}}\geq \frac{\DSE{\Theta}-I_0-1}{\log\abs{\vec{\Theta}}},\nonumber
		\end{align} 
		where $\DSE{\Theta}$ is the discrete entropy of $\Theta$, $I_0$ is the leakage level of private information in \eqref{Eq: OP} and $\abs{\vec{\Theta}}$ denotes the cardinality of the set $\vec{\Theta}$.
	\end{theorem}
	\begin{IEEEproof}
		See Appendix \ref{App: EP-LB}.
		\end{IEEEproof}
	According to Theorem \ref{Theo: EP-LB}, the error probability of any estimator of the private parameter based on the output of the privacy filter is limited by the leakage level of private information $I_0$. The lower bound in Theorem \eqref{Theo: EP-LB} increases as the leakage level of private information becomes small. Thus, for a relatively small value of $I_0$, no estimator can reliably infer the private parameter $\Theta$ which implies that the proposed framework is capable of ensuring the privacy of $\Theta$. 
	     
	In Appendix \ref{App: EP-LB}, we show that the mutual information between $\Theta$ and $\tilde{Y}_{1:T}$ can be upper bounded as 
	\begin{align}\label{Eq: MI-UB-T}
	\MI{\Theta}{\tilde{Y}_{1:T}} \leq  \MI{\Theta}{\tilde{\Theta}}.
	\end{align}
	Note that $\MI{\Theta}{\tilde{Y}_{1:T}}$ quantifies the amount of information that can be inferred about $\Theta$ by observing $\tilde{Y}_{1:T}$. Thus, the inequality above implies that the leakage of information about the private parameter via the output of the privacy filter is limited by the mutual information between the input and the output of the randomizer. Hence, the upper bound on the mutual information between $\Theta$ and $\tilde{\Theta}$ in \eqref{Eq: OP} ensures the privacy of $\Theta$. 
	\begin{remark}
		\textcolor{black}{To prove Theorem \ref{Theo: EP-LB}, we first show that the Markov chain $\Theta\longrightarrow\tilde{\Theta}\longrightarrow \tilde{Y}_{1:T}$ holds (see Appendix \ref{App: Dist_Gen} for more details). This Markov chain along with the data processing inequality \cite{CT06} allow us to establish the inequality in \eqref{Eq: MI-UB-T}. Finally, the lower bound in Theorem \ref{Theo: EP-LB} is obtained by combining Fano's inequality with the inequality \eqref{Eq: MI-UB-T}.}
	\end{remark}
	\section{Special Cases}\label{Sec: SC}
	In section \ref{Sec: MR}, we studied the structure of the nonlinear transformation without imposing any restriction on the joint probability density function (p.d.f.) of the measurements. In this section, we study the structure of the nonlinear transformation in the special cases of independent and identically distributed (i.i.d.), Markovian and Gauss-Markov measurements. 
	\subsection{Independent and Identically Distributed Measurements}
	In this subsection, we assume that $Y_1,\dots,Y_T$ is a sequence of independent and identically distributed (i.i.d.) random variables with the common p.d.f. $p_{\theta_i}\paren{y^1,\dots,y^d}$ when $\Theta$ is equal to $\theta_i$. Given $\Theta=\theta_i$ and $\tilde{\Theta}=\tilde{\theta}_j$, the privacy filter at time-step $k$ generates $y^l_k$ according to 
	\begin{align}
	\tilde{y}^l_k&=\INVCCDF{{l,\tilde{\theta}_j}}{u^l_k}{\tilde{y}^{1:l-1}_k},\nonumber\\
	u^l_k&=\CCDF{l,\theta_i}{y^l_k}{{y}_k^{1:l-1}},\nonumber
	\end{align}
	where 
	\begin{align}
		\CCDF{l,\theta_i}{z}{y^{1:l-1}_k}&=\int_{-\infty}^{z}p_{\theta_i}\paren{y^l\left|y^{1:l-1}_k\right.}dy^l.\nonumber
		\end{align}
	Note that 	$\tilde{Y}_k$ is generated only using $Y_k$ when the measurements are independent and identically distributed.
	\subsection{Markovian Measurements}
	When measurements are Markovian,  the conditional joint p.d.f. of $Y_1,\dots,Y_T$  given $\Theta=\theta_i$ factorizes as 
	\begin{align}
\textcolor{black}{	p_{\theta_i}\paren{y_1,\dots,y_T}=\DF{1,\theta_i}{y_1}\prod_{k=1}^{T-1}\CD{k+1,\theta_i}{y_{k+1}}{y_k}.}\nonumber
	\end{align}
In this case, the privacy filter at time-step $k$ generates $\tilde{y}^l_k$ according to 
	\begin{align}
	\tilde{y}^l_k&=\INVCCDF{{l,k,\tilde{\theta}_j}}{u^l_k}{\tilde{y}^{1:l-1}_k,\tilde{y}_{k-1}},\nonumber\\
	u^l_k&=\CCDF{l,k,\theta_i}{y^l_k}{{y}_k^{1:l-1},y_{k-1}},\nonumber	
	\end{align}
	where 
	\begin{align}
	\textcolor{black}{\CCDF{l,k,\theta_i}{z}{y^{1:l-1}_k,y_{k-1}}}&=\int_{-\infty}^{z}p_{k,\theta_i}\paren{y^l\left|y^{1:l-1}_k,y_{k-1}\right.}dy^l.\nonumber
	\end{align}
	Note that, different from the general case, $\tilde{Y}_k$ is generated using $Y_{k-1}$ and $Y_{k}$ in the Markovian case. 
	\subsection{Gauss-Markov Measurements}\label{Subsec: GM}
	Given $\Theta=\theta_i$ in the Gauss-Markov case, the measurements are generated by the following model:
	\begin{align}\label{Eq: Model-GM-1}
	X_{k+1}&=A_iX_k+W_k\nonumber\\
	Y_k&=C_iX_k+V_k,
	\end{align}
	where $\left\{W_k\right\}_k$ and $\left\{V_k\right\}_k$ are sequences of i.i.d. zero mean Gaussian random vectors with the covariance matrices $Q^i_w$ and $Q^i_v$, respectively. We assume that $X_0$ is a zero mean Gaussian random vector with the covariance matrix $Q^i_0$. We also assume $\left\{W_k\right\}_k$ and $\left\{V_k\right\}_k$ are mutually independent and $X_0$ is independent of $\left\{W_k,V_k\right\}_k$. For all $i$, we assume that $A_i$ is Schur stable,  $\paren{A_i,C_i}$ is observable and the matrices $Q^i_w,Q^i_v,Q^i_0$ are positive definite.   

	 In the Gauss-Markov case, the parameter $\theta_i=\paren{A_i,C_i,Q^i_w,Q^i_v,Q^i_0}$ characterizes the joint p.d.f. of the measurements.	Given $\tilde{\Theta}=\tilde{\theta}_j=\paren{A_j,C_j,Q^j_w,Q^j_v,Q^j_0}$, the objective of the nonlinear transformation is to sequentially generate $\tilde{Y}_1,\dots,\tilde{Y}_T$ such that their joint p.d.f.  is the same as the joint p.d.f. of  a sequence generated by a Gauss-Markov model with the parameters $\paren{A_j,C_j,Q^j_w,Q^j_v,Q^j_0}$.
			
		Before proceeding with the structure of the nonlinear transformation in the Gauss-Markov case, we introduce the structure of the optimal output predictor of the model in \eqref{Eq: Model-GM-1}.	Let $\hat{x}_{k\left|k\right.}$ denote the Kalman estimate of $X_k$ based on $\left\{ Y_1=y_1,\dots,Y_{k}=y_{k},\Theta=\theta_i\right\}$ which is given by \eqref{Eq: KE-i} with $\hat{x}_{1\left|0\right.}=0$ where $\Sigma^i_{k\left|k-1\right.}$ is the one-step ahead prediction error covariance which satisfies the algebraic Riccati recursion in  \eqref{Eq: RR-i}.
		\begin{figure*}
			\begin{align}\label{Eq: KE-i}
			\hat{x}_{k\left|k\right.}&=\hat{x}_{k\left|k-1\right.}+\Sigma^i_{k\left|k-1\right.}C_i^\top\paren{C_i\Sigma^i_{k\left|k-1\right.}C_i^\top+Q^i_v}^{-1}\paren{y_k-C_i\hat{x}_{k\left|k-1\right.}}.\nonumber\\
			\hat{x}_{k\left|k-1\right.}&=A_i\hat{x}_{k-1\left|k-1\right.}.			
			\end{align}
			\hrule
			\begin{align}\label{Eq: RR-i}
			\Sigma^i_{k+1\left|k\right.}=A_i\Sigma^i_{k\left|k-1\right.}A_i^\top+Q^i_w-A_i\Sigma^i_{k\left|k-1\right.}C_i^\top\paren{C_i\Sigma^i_{k\left|k-1\right.}C_i^\top+Q^i_v}^{-1}C_i\Sigma^i_{k\left|k-1\right.}A_i^\top.
			\end{align}
			\hrule
			\begin{align}\label{Eq: KE-j}
			\hat{\tilde{y}}_{k\left|k-1\right.}&=C_j\hat{\tilde{x}}_{k\left|k-1\right.},\nonumber\\
			\hat{\tilde{x}}_{k\left|k\right.}&=\hat{\tilde{x}}_{k\left|k-1\right.}+\Sigma^j_{k\left|k-1\right.}C_j^\top\paren{C_j\Sigma^j_{k\left|k-1\right.}C_j^\top+Q^j_v}^{-1}\paren{\tilde{y}_k-C_j\hat{\tilde{x}}_{k\left|k-1\right.}}.\nonumber\\
				\hat{\tilde{x}}_{k\left|k-1\right.}&=A_j\hat{\tilde{x}}_{k-1\left|k-1\right.}.
			\end{align}
			\hrule	
			\begin{align}\label{Eq: RR-j}
			\Sigma^j_{k+1\left|k\right.}=A_j\Sigma^j_{k\left|k-1\right.}A_j^\top+Q^j_w-A_j\Sigma^j_{k\left|k-1\right.}C_j^\top\paren{C_j\Sigma^j_{k\left|k-1\right.}C_j^\top+Q^j_v}^{-1}C_j\Sigma^j_{k\left|k-1\right.}A_j^\top.
			\end{align}
			\hrule		
		\end{figure*}
	Let $\hat{y}_{k\left|k-1\right.}$ denote the  one-step ahead Kalman predictor of $Y_k$ based on $\left\{ Y_1=y_1,\dots,Y_{k-1}=y_{k-1},\Theta=\theta_i\right\}$ which is given  by 
		\begin{align}
		\hat{y}_{k\left|k-1\right.}=C_i\hat{x}_{k\left|k-1\right.},\nonumber 
		\end{align}
		with the output prediction error covariance matrix  $\Sigma^{o,i}_{k\left|k-1\right.}$ defined as 
		\begin{align}
		\Sigma^{o,i}_{k\left|k-1\right.}=C_i\Sigma^i_{k\left|k-1\right.}C_i^\top+Q^i_v.\nonumber 
		\end{align}
		
\begin{figure*}
	\centering{\includegraphics[scale=0.5]{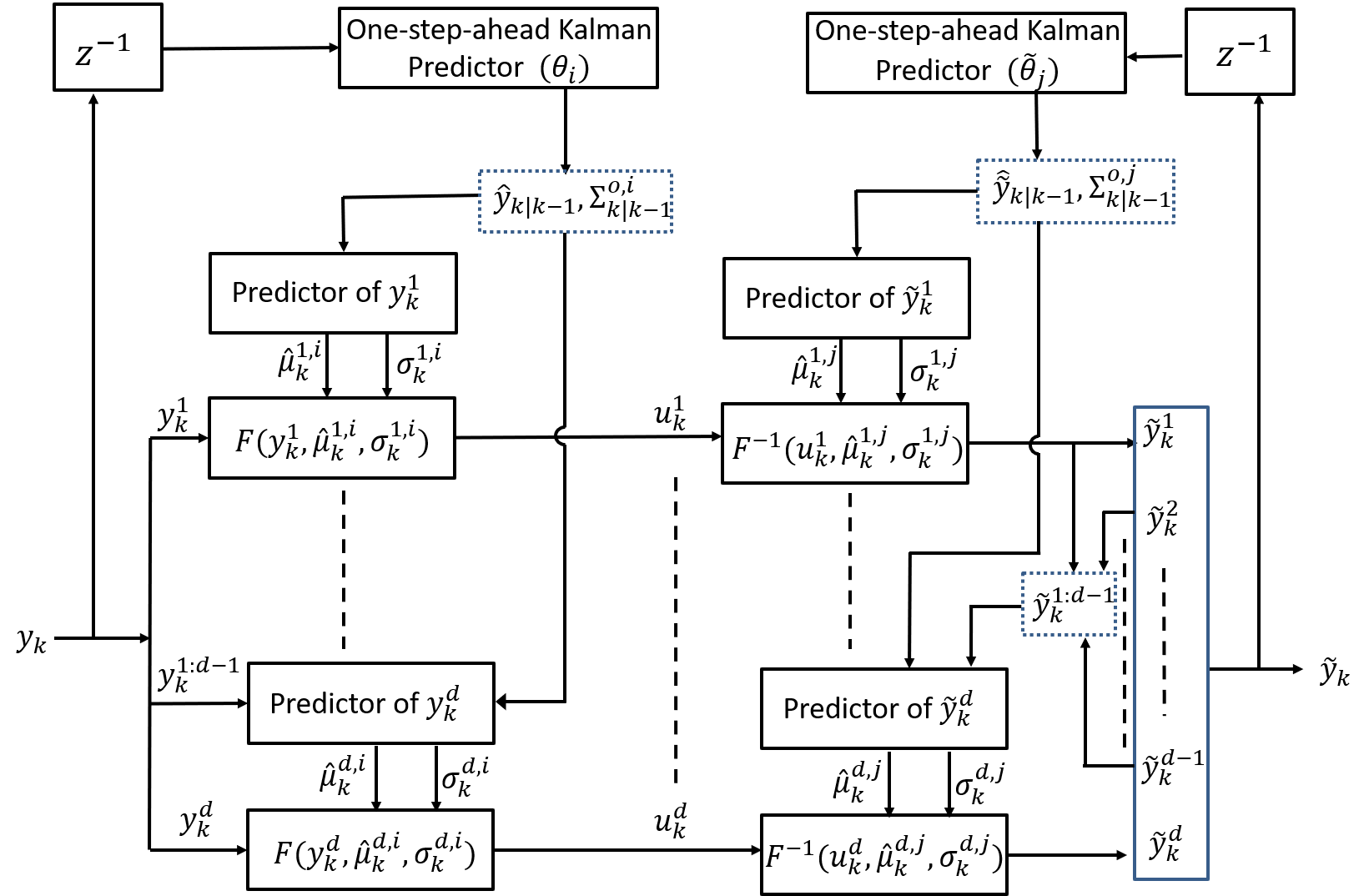}}
	\caption{The structure of the nonlinear transformation in the Gauss-Markov case.}\label{Fig: GM}
\end{figure*}

		The next lemma studies the p.d.f. of the $l$th component of $Y_k$ given \newline \textcolor{black}{$\left\{Y^{1:l-1}_k=y^{1:l-1}_k,Y_{1:k-1}=y_{1:k-1},\Theta=\theta_i\right\}$}. This result will be used to study the structure of the nonlinear transformation in the Gauss-Markov case. 
		\begin{lemma}\label{Lem: OCAP}
			The conditional p.d.f. of $Y^l_k$ given $\left\{Y^{1:l-1}_k=y^{1:l-1}_k,Y_{1:k-1}=y_{1:k-1},\Theta=\theta_i\right\}$ is a Gaussian p.d.f. with mean $\hat{\mu}^{li}_{k}$ and variance $\sigma^{l,i}_{k}$ given by  
			\begin{align}\label{Eq: MV}
			\hat{\mu}^{l,i}_k&=\hat{y}^l_{k\left|k-1\right.}-\Delta^{l,i}_k\left[\Sigma^{o,i}_{k\left|k-1\right.}\right]^{-1}_{l-1}\paren{y^{1:l-1}_k-\hat{y}^{1:l-1}_{k\left|k-1\right.}}\nonumber\\
			\sigma^{l,i}_{k}&=\Sigma^{l,o,i}_{k\left|k-1\right.}-\Delta^{l,i}_k\left[\Sigma^{o,i}_{k\left|k-1\right.}\right]^{-1}_{l-1}\paren{\Delta^{l,i}_k}^\top.
			\end{align}
			where $\hat{y}^l_{k\left|k-1\right.}$ is the $l$th element of $\hat{y}_{k\left|k-1\right.}$, $\hat{y}^{1:l-1}_{k\left|k-1\right.}$ is the vector of the first $l-1$ elements of   $\hat{y}_{k\left|k-1\right.}$,      $\left[\Sigma^{o,i}_{k\left|k-1\right.}\right]_{l-1}$ is a matrix formed by the elements in the first $l-1$ rows and the first $l-1$ columns of $\Sigma^{o,i}_{k\left|k-1\right.}$ with $\left[\Sigma^{o,i}_{k\left|k-1\right.}\right]_{0}^{-1}=0$,  $\Sigma^{l,o,i}_{k\left|k-1\right.}$ is the $l$th diagonal entry of $\Sigma^{o,i}_{k\left|k-1\right.}$ and $\Delta^{l,i}_k$ is the vector of the first $l-1$ elements in the $l$th row of $\Sigma^{o,i}_{k\left|k-1\right.}$.
		\end{lemma}
	\begin{IEEEproof}
		See Appendix \ref{App: GD}.
		\end{IEEEproof}
	Lemma \ref{Lem: OCAP} implies that $\hat{\mu}^{l,i}_k$ is the minimum mean square error (MMSE) predictor of $Y_k$ given \textcolor{black}{$\left\{Y^{1:l-1}_k=y^{1:l-1}_k,Y_{1:k-1}=y_{1:k-1},\Theta=\theta_i\right\}$} and $\sigma^{l,i}_{k}$ is the error variance associated with $	\hat{\mu}^{l,i}_{k}$. The next theorem studies the structure of the nonlinear transformation in the Gauss-Markov case. 
		\begin{theorem}\label{Theo: GM}
		Consider the Gauss-Markov model with $\Theta=\theta_i$ and $\tilde{\Theta}=\tilde{\theta}_j$. Then, for $1\leq l\leq d$, the $l$th component of the $\tilde{Y}^l_k$ is generated according to   
		\begin{align}
			\tilde{y}^l_k&=F^{-1}\paren{u^l_k,\hat{\mu}^{l,j}_{k},\sigma^{l,j}_{k}}\nonumber\\
		u^l_k&=F\paren{y^l_k,\hat{\mu}^{l,i}_k,\sigma^{l,i}_{k}},\nonumber	
		\end{align}
		where $F\paren{\cdot, \hat{\mu},\sigma}$ is the cumulative distribution function of a Gaussian random variable with mean $\hat{\mu}$ and variance $\sigma$, $F^{-1}\paren{\cdot, \hat{\mu},\sigma}$ is the inverse function of $F\paren{\cdot, \hat{\mu},\sigma}$,  $\hat{\tilde{y}}_{k\left|k-1\right.}$ and $\Sigma^{o,j}_{k\left|k-1\right.}$ are given by \eqref{Eq: KE-j} and \eqref{Eq: RR-j} with $\hat{\tilde{x}}_{1\left|0\right.}=0$, $\hat{\mu}^{l,i}_k$ and  $\sigma^{l,i}_{k}$ are defined in \eqref{Eq: MV}, and $\sigma^{l,j}_{k}$ and $\hat{\mu}^{l,j}_{k}$ are defined in a similar way. 
        
	\end{theorem} 
\begin{IEEEproof}
\textcolor{black}{According to Lemma \ref{Lem: OCAP}, the conditional p.d.f. of $Y^l_k$ given \newline $\left\{Y^{1:l-1}_k=y^{1:l-1}_k,Y_{1:k-1}=y_{1:k-1},\Theta=\theta_i\right\}$ is a Gaussian p.d.f. with mean $\hat{\mu}^{li}_{k}$ and variance $\sigma^{l,i}_{k}$. Thus, we have}
\begin{align}
p_{\theta_i}\paren{x\left|y^{1:l-1}_k,y_{1:k-1}\right.}=\frac{1}{\sqrt{2\pi\sigma^{l,i}_{k} }}\e{-\frac{1}{2\sigma^{l,i}_{k}}\paren{x-\hat{\mu}^{li}_{k}}^2}.\nonumber
\end{align}
\textcolor{black}{This implies that $\CCDF{l,k,\theta_i}{z}{y^{1:l-1}_k,y_{1:k-1}}=F\paren{z,\hat{\mu}^{l,i}_k,\sigma^{l,i}_{k}}$ where $F\paren{\cdot, \hat{\mu},\sigma}$ is the cumulative distribution function of a Gaussian random variable with mean $\hat{\mu}$ and variance $\sigma$. Using a similar argument, it is straightforward to show $\INVCCDF{{l,k,\tilde{\theta}_j}}{z}{\tilde{y}^{1:l-1}_k,\tilde{y}_{1:k-1}}= F^{-1}\paren{z,\hat{\mu}^{l,j}_{k},\sigma^{l,j}_{k}}$ where $F^{-1}\paren{\cdot, \hat{\mu},\sigma}$ is the inverse function of $F\paren{\cdot, \hat{\mu},\sigma}$. Using  these observations and Theorem \ref{Theo: Dist_Gen}, $\tilde{y}^l_k$, in the Gauss-Markov case, is generated according to }
	\begin{align}
\tilde{y}^l_k&=F^{-1}\paren{u^l_k,\hat{\mu}^{l,j}_{k},\sigma^{l,j}_{k}},\nonumber\\
u^l_k&=F\paren{y^l_k,\hat{\mu}^{l,i}_k,\sigma^{l,i}_{k}},\nonumber	
\end{align}
which completes the proof.
\end{IEEEproof}
According to Theorem \ref{Theo: GM}, the structure of the nonlinear transformation in the Gauss-Markov case is characterized by two types of predictors: one-step ahead Kalman predictors and one-component ahead predictors. At time-step $k$, the Kalman predictors compute $\hat{y}_{k\left|k-1\right.}$ and $\hat{\tilde{y}}_{k\left|k-1\right.}$ which, respectively, are the optimal prediction of $Y_k$ given $\left\{Y_{1:k-1}=y_{1:k-1},\Theta=\theta_i\right\}$ and the optimal prediction of $\tilde{Y}_k$ given $\left\{\tilde{Y}_{1:k-1}=\tilde{y}_{1:k-1},\tilde{\Theta}=\tilde{\theta}_j\right\}$. At time-step $k$, the optimal one-component ahead predictors use $\hat{y}_{k\left|k-1\right.}$ and $\hat{\tilde{y}}_{k\left|k-1\right.}$  to compute $\hat{\mu}^{l,i}$ and $\hat{\mu}^{l,j}$ which are the MMSE prediction of ${Y}^l_k$  given  $\left\{{Y}^{1:l-1}_{k}={y}^{1:l-1}_{k},{Y}_{1:k-1}={y}_{1:k-1},{\Theta}=\theta_i\right\}$ and the MMSE prediction of $\tilde{Y}^l_k$ given $\left\{\tilde{Y}^{1:l-1}_{k}=\tilde{y}^{1:l-1}_{k},\tilde{Y}_{1:k-1}=\tilde{y}_{1:k-1},\tilde{\Theta}=\tilde{\theta}_j\right\}$, respectively. The outputs of these predictors are used to compute the parameters of the nonlinear transformation in the Gauss-Markov case as shown in Fig. \ref{Fig: GM}. 
\begin{remark}
	\textcolor{black}{The Kalman predictors significantly reduce the computational complexity of generating the disguised measurements. In the Gauss-Markov case, the proposed filtering scheme requires the conditional p.d.f.s $p_{\theta_i}\paren{y_k\left|y_{1:k-1}\right.}$ and $p_{\theta_i}\paren{\tilde{y}_k\left|\tilde{y}_{1:k-1}\right.}$ which can be computed (recursively) with low computational costs using the filtering equations in \eqref{Eq: KE-i}-\eqref{Eq: RR-j}. Note that the direct computation of these p.d.f.s becomes prohibitive when $k$ is large.}
\end{remark} 
	\section{Numerical Results}\label{Sec: NR}
	In this section, we numerically evaluate the performance of the proposed framework in ensuring the occupancy privacy of the $\rm CO_{2}$ measurements in a building automation application. To this end, let $Y_k$ denote the  $\rm CO_{2}$ measurement inside a room at time-step $k$ which evolves according to 	
		\begin{align}
	{X}_{k+1}&=0.95{X}_k+{W}_k+10\Theta,\nonumber\\
	{Y}_{k}&={X}_k+{V}_k,\nonumber
	\end{align}
	where $\Theta$ denotes the number of occupants in the room, $\left\{W_k,V_k\right\}_k$ is a sequence of i.i.d. Gaussian random variables with zero mean and variance $10^{-1}$ and $X_0$ is a Gaussian random variable, independent of $\left\{W_k,V_k\right\}_k$, with mean $100$ and variance $1$. The occupancy parameter $\Theta$ is assumed to take values in $\left\{0,1\right\}$ with the probabilities $\PRP{\Theta=1}=\PRP{\Theta=0}=0.5$. We also assume that the pseudo occupancy parameter $\tilde{\Theta}$ takes values in $\left\{0,0.2,0.4,0.6,0.8,1\right\}$. The horizon length $T$ was set to $50$ in our simulations.  
	
\begin{figure}
		\begin{center}
		\includegraphics[scale=0.6]{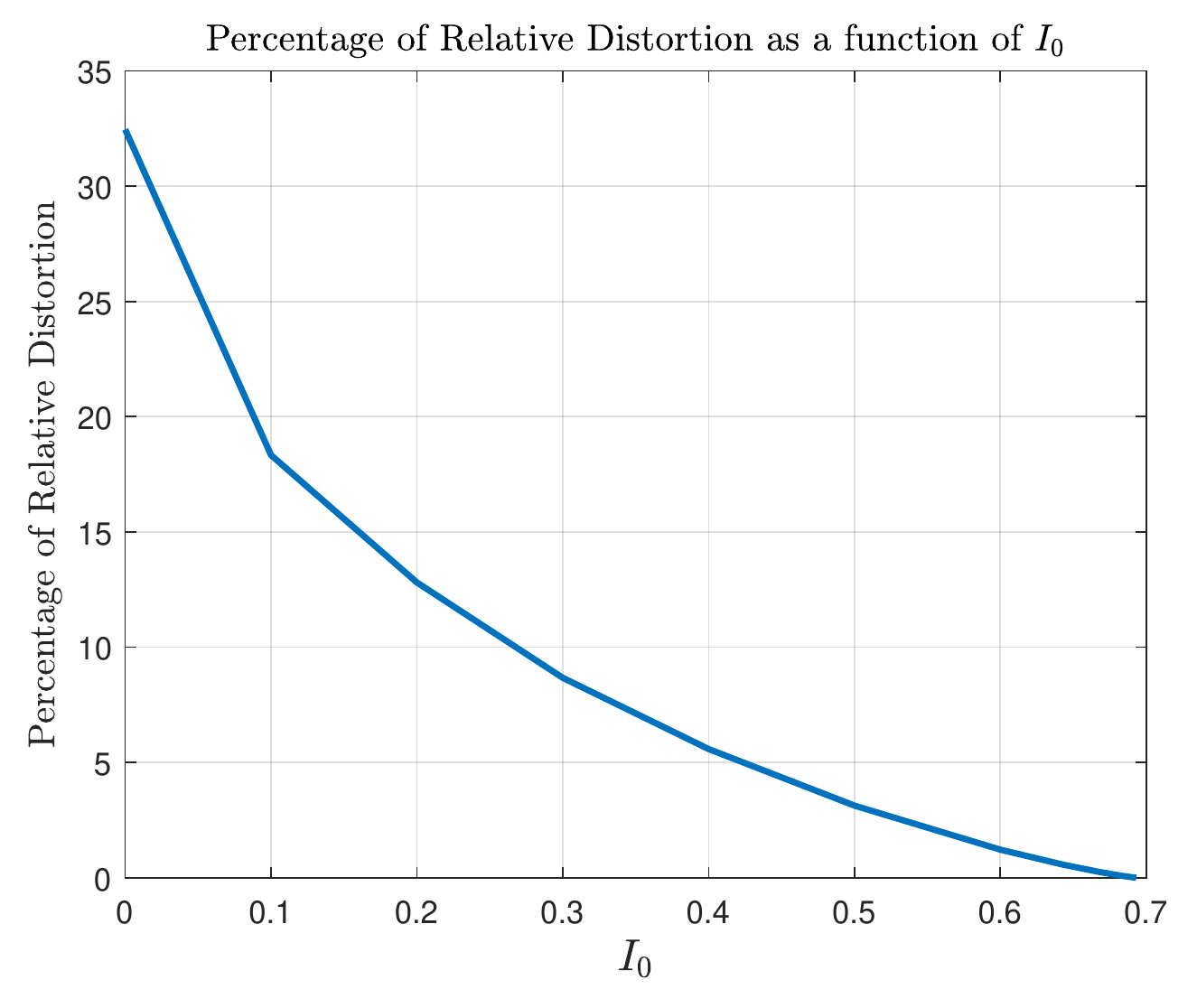}
				\end{center}
		\caption{The percentage of relative distortion versus the leakage level of private information $I_0$.}\label{Fig: Dist}
	\end{figure}

Fig. \ref{Fig: Dist} shows the percentage of the relative distortion, due to the privacy filter, as a function of the leakage level of private information $I_0$. The relative distortion is defined as the ratio of the average distortion between the input and output of the filter over the average $\rm CO_2$. According to Fig. \ref{Fig: Dist}, the relative distortion increases as the leakage of private information becomes small since the privacy constraint in \eqref{Eq: OP} becomes tight in this case. The maximum distortion and maximum privacy level are achieved when leakage of private information is equal to zero. Note that $\Theta$ and $\tilde{\Theta}$ are independent when $I_0=0$ which results in a high level of relative distortion. The minimum distortion and minimum privacy levels are achieved for $I_0=0.69$. In this case, the privacy constraint is relaxed and however $\Theta$ can be correctly inferred from $\tilde{\Theta}$.

To study the performance of an adversary in estimating the occupancy using $\tilde{Y}_k$, we consider the following estimator of the occupancy 
\begin{align}
\hat{\Theta}\paren{\tilde{Y}}=\arg\min_{\theta\in\left\{0,1\right\}}\abs{\frac{\bar{Y}-\bar{\bar{Y}}}{0.95}
-\theta},\nonumber
\end{align}
where $\bar{Y}$ and $\bar{\bar{Y}}$ are given by 
	\begin{align}
\bar{Y}&=\frac{1}{10}\sum_{i=41}^{50}\tilde{Y}_i, \nonumber\\
\bar{\bar{Y}}&=\frac{1}{10}\sum_{i=40}^{49}\tilde{Y}_i.\nonumber
\end{align}
 
\begin{figure}
				\begin{center}
	\includegraphics[scale=0.6]{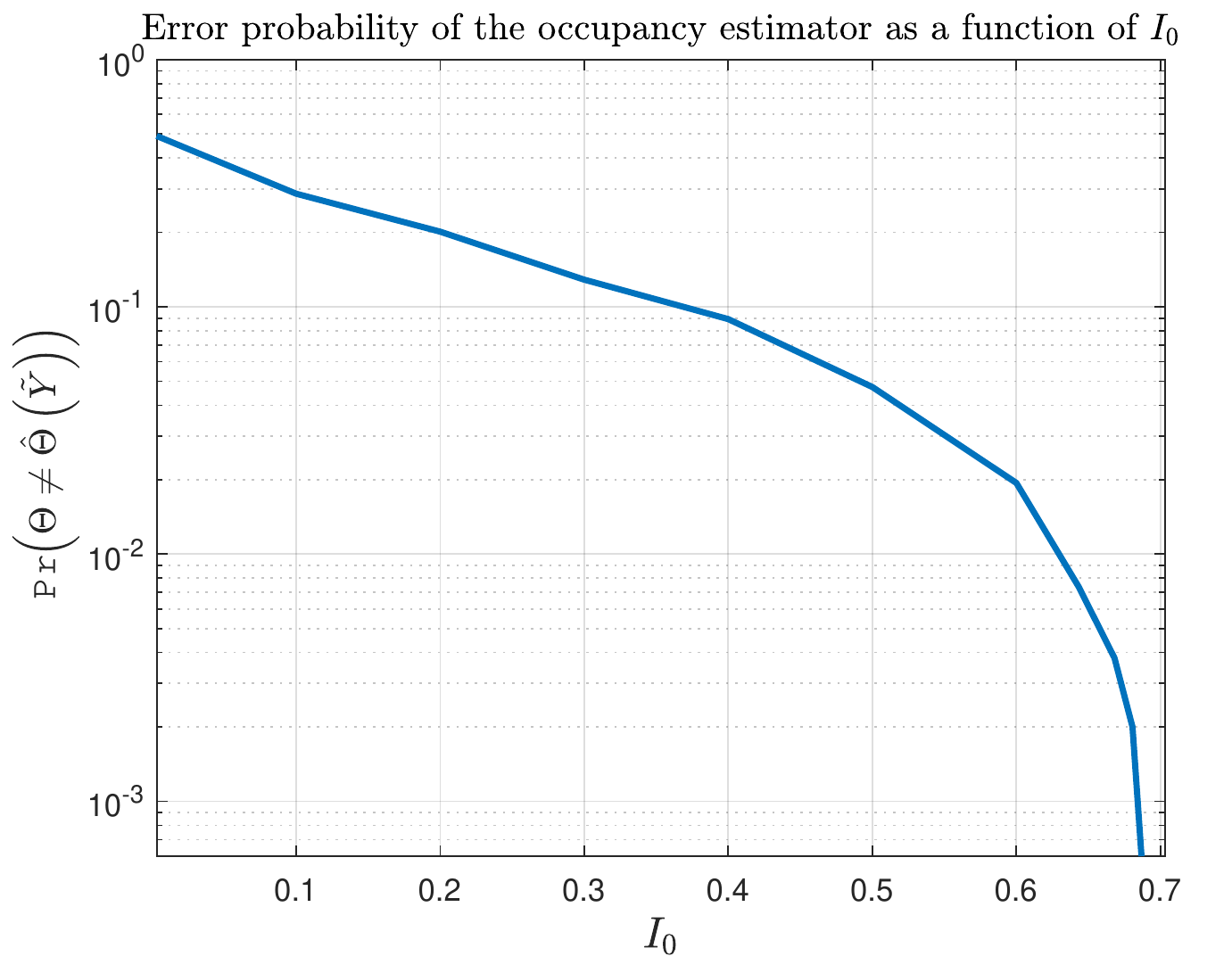}
			\end{center}
	\caption{The error probability of the occupancy estimator versus the leakage level of private information $I_0$.}\label{Fig: EP}
\end{figure}

 Fig. \ref{Fig: EP} shows the error probability of the proposed occupancy estimator as a function of the leakage level of private information $I_0$. Based on this figure, the proposed estimator can reliably estimate the occupancy information when the leakage of private information is high. However, as $I_0$ decreases from $0.68$ to $0.3$, the performance of the occupancy estimator degrades by more than two orders of magnitude while the distortion due to the privacy filter at $I_0=0.3$ is approximately $0.8$ percent. This is due to the fact that the output of the filter ceases to be a reliable source of information for estimating the occupancy as the leakage of private information decreases.  

	\begin{figure}
		\begin{center}
				\includegraphics[scale=0.6]{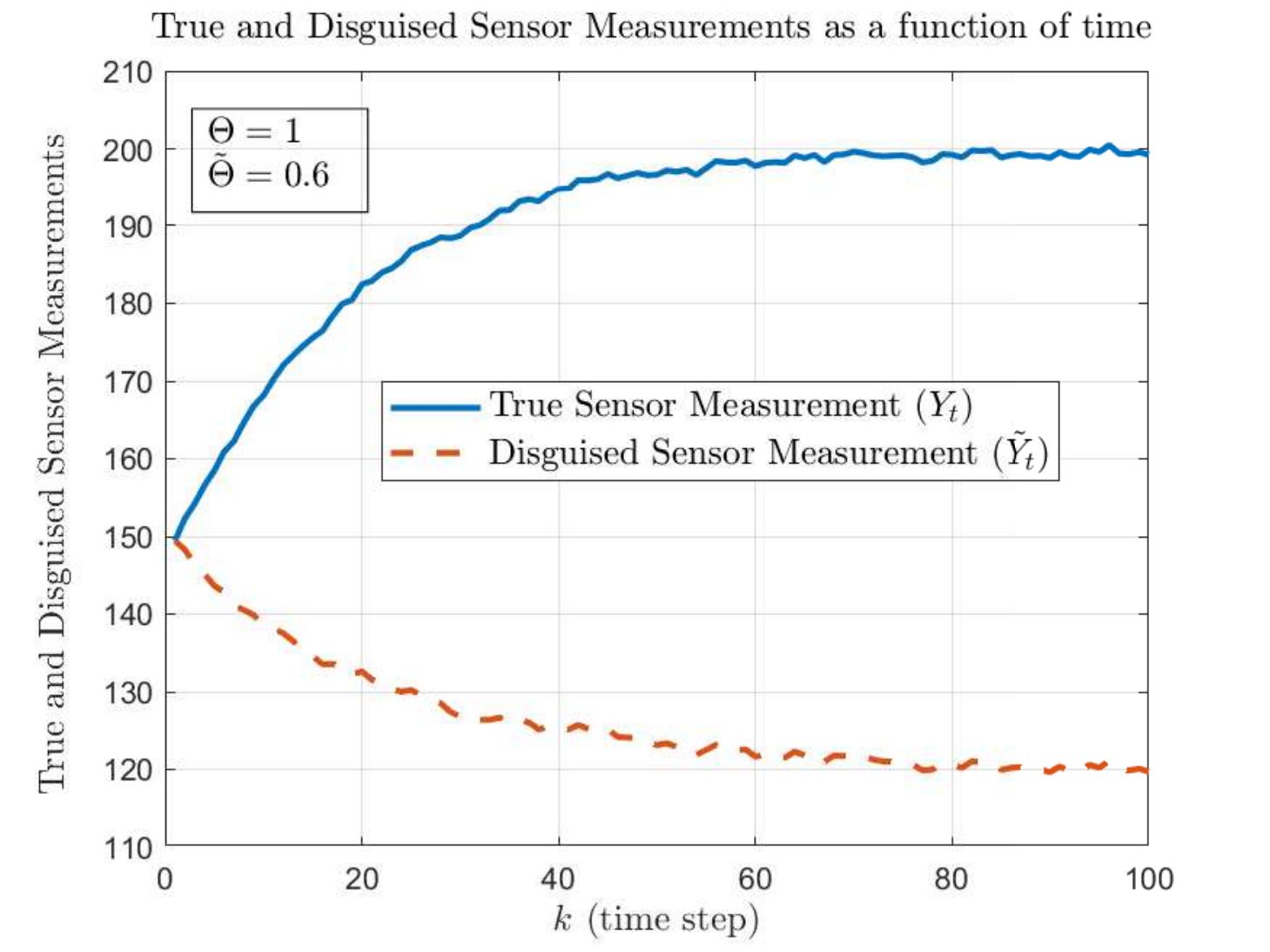}
		\end{center}

	\caption{\textcolor{black}{Realizations of the true and disguised sensor measurements as a function of time for $\Theta=1$, and $\tilde{\Theta}=0.6$}.}\label{Fig: TD}
\end{figure}

	\begin{figure}
			\begin{center}
	\includegraphics[scale=0.6]{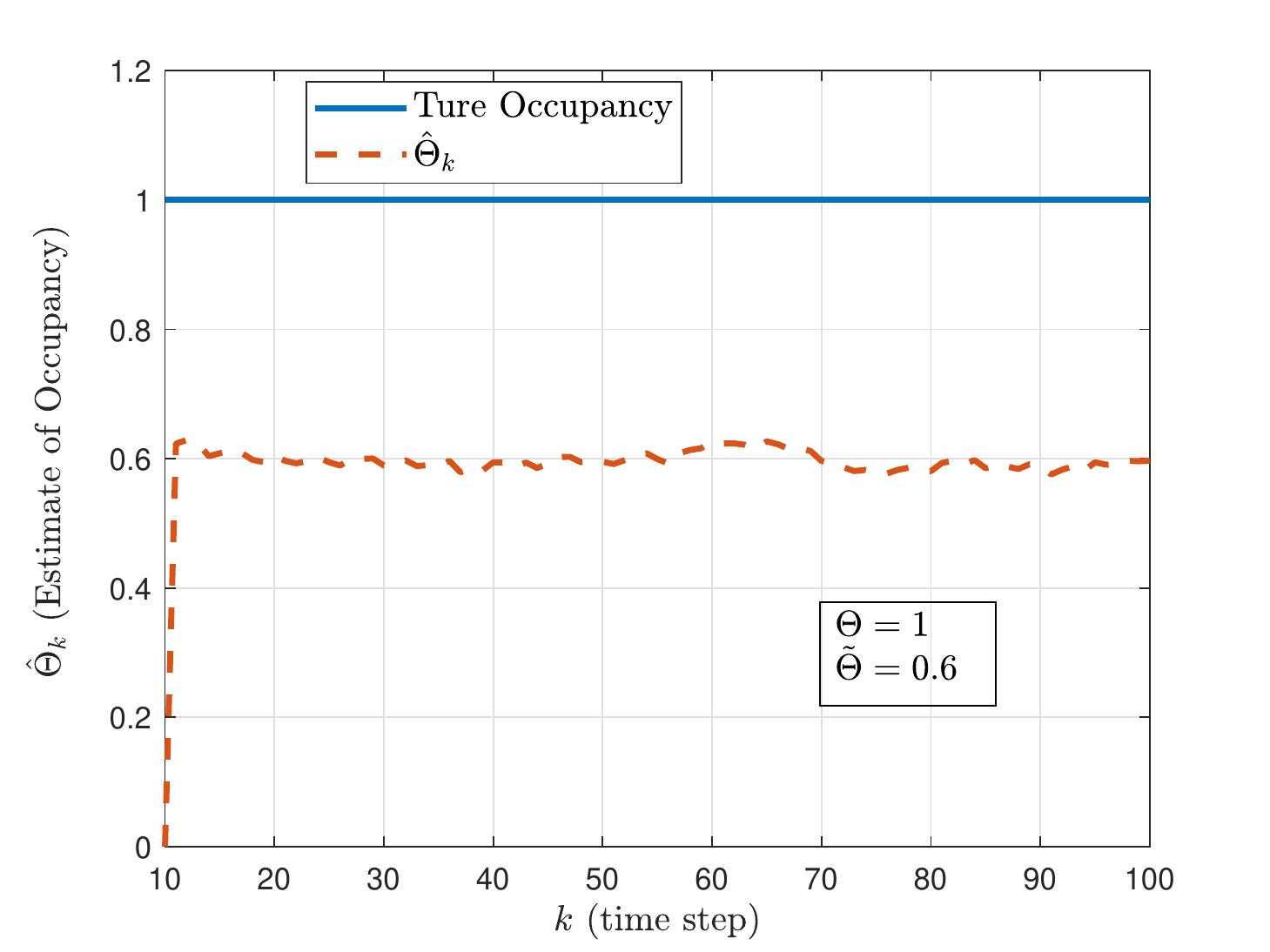}
			\end{center}
	\caption{\textcolor{black}{The output of the occupancy estimator, based on the disguised measurements, as a function of time}.}\label{Fig: ESR}
\end{figure}

\textcolor{black}{Fig. \ref{Fig: TD} shows realizations of the true and disguised sensor measurements as a function of time for $\Theta=1$ and $\tilde{\Theta}=0.6$. According to this figure, the privacy filter results in a certain level of distortion between the true and disguised measurements. Fig. \ref{Fig: ESR} shows the output of the occupancy estimator in \eqref{Eq: O-E} when the output of the nonlinear transformation is used as the input to the estimator. According to this figure, the occupancy estimator cannot accurately infer the occupancy based on the disguised measurements.  }
\section{Conclusions}\label{Sec: Conc}
In this paper, we proposed a privacy filter design framework to ensure the statistical parameter privacy of a sequence of sensor measurements. Under the proposed framework, the privacy filter has two components: a randomizer and a nonlinear transformation. The optimal design of randomizer was studied under a privacy constraint and it was shown than the optimal randomizer is the solution of a convex optimization problem. The privacy level of the proposed framework was examined using information-theoretic inequalities.  We also studied the structure of the nonlinear transformation in special cases.   

\textcolor{black}{ An important direction for our future research is to investigate the optimal design of the randomization probabilities and nonlinear transformation for the privacy-aware closed-loop control problem. The optimal design of the randomizer when the private parameter is time-varying and the randomizer has the causal knowledge of the private parameter is another important avenue for our future work.  }
	\appendices
	\section{Proof of Theorem \ref{Theo: Dist_Gen}}\label{App: Dist_Gen}
	To prove this result, we first show that the Markov chain $\Theta\longrightarrow\tilde{\Theta}\longrightarrow \tilde{Y}_{1:T}$ holds. To this end, we derive an expression for the conditional p.d.f. of $\tilde{Y}^l_k$ given the event \newline $\left\{\tilde{Y}_k^{1:l-1}=\tilde{y}^{1:l-1}_k,\tilde{Y}_{1:k-1}=\tilde{y}_{1:k-1},\Theta=\theta_i, \tilde{\Theta}=\tilde{\theta}_j\right\}$ in the next lemma. 
	
	\begin{lemma}\label{Lem: Density}
		Let $\CPDF{\tilde{Y}^l_{k}}{\tilde{y}^l_{k}}{\tilde{y}^{1:l-1}_k,\tilde{y}_{1:k-1},\Theta=\theta_i,\tilde{\Theta}=\tilde{\theta}_j}$ denote the conditional p.d.f. of $\tilde{Y}^l_k$ given the event $\left\{\tilde{Y}_k^{1:l-1}=\tilde{y}^{1:l-1}_k,\tilde{Y}_{1:k-1}=\tilde{y}_{1:k-1},\Theta=\theta_i, \tilde{\Theta}=\tilde{\theta}_j\right\}$. Then, we have 
		\begin{align}
		&\CPDF{\tilde{Y}^l_{k}}{\tilde{y}^l_{k}}{\tilde{y}^{1:l-1}_k,\tilde{y}_{1:k-1},\Theta=\theta_i,\tilde{\Theta}=\tilde{\theta}_j}\nonumber\\
		&\hspace{4cm}=\CPDF{\tilde{\theta}_j}{\tilde{y}^l_k}{\tilde{y}^{1:l-1}_k,\tilde{y}_{1:k-1}},\nonumber
		\end{align}
		where $\CPDF{\tilde{\theta}_j}{\tilde{y}^l_k}{\tilde{y}^{1:l-1}_k,\tilde{y}_{1:k-1}}$ is obtained from $p_{\tilde{\theta}_j}\paren{\tilde{y}_1,\dots,\tilde{y}_T}$ using the Bayes' rule and marginalization. 
	\end{lemma}
	\begin{IEEEproof}
		See Appendix \ref{App: Density}
	\end{IEEEproof}

	Let $\CPDF{\tilde{Y}_{1:T}}{\tilde{y}_{1:T}}{\Theta=\theta_i,\tilde{\Theta}=\tilde{\theta}_j}$ denote the  conditional p.d.f. of $\tilde{Y}_{1:T}$ given $\left\{\Theta=\theta_i,\tilde{\Theta}=\tilde{\theta}_j\right\}$ which can be written as \eqref{Eq: BR}
	 \begin{figure*}
		\begin{align}\label{Eq: BR}
		\CPDF{\tilde{Y}_{1:T}}{\tilde{y}_{1:T}}{\Theta=\theta_i,\tilde{\Theta}=\tilde{\theta}_j}&\stackrel{(a)}{=}\prod_{k=1}^T\prod_{l=1}^d\CPDF{\tilde{Y}^l_{k}}{\tilde{y}^l_{k}}{\tilde{y}^{1:l-1}_k,\tilde{y}_{1:k-1},\Theta=\theta_i,\tilde{\Theta}=\tilde{\theta}_j}\nonumber\\
		&\stackrel{(b)}{=}\prod_{k=1}^T\prod_{l=1}^d \CD{\tilde{\theta}_j}{\tilde{y}^l}{\tilde{y}^{1:l-1}_k,\tilde{y}_{1:k-1}}\nonumber\\
		&\stackrel{(c)}{=}p_{\tilde{\theta}_j}\paren{\tilde{y}_1,\dots,\tilde{y}_T}.
		\end{align}
		\hrule
	\end{figure*}
	where $(a)$ and $(c)$ follow from the Bayes' rule and $(b)$ follows from Lemma \ref{Lem: Density}. This implies that the conditional p.d.f. of  $\tilde{Y}_1,\dots,\tilde{Y}_T$ given $\Theta$ and $\tilde{\Theta}$ only depends on $\tilde{\Theta}$. Thus, the following Markov chain holds $\Theta\longrightarrow\tilde{\Theta}\longrightarrow \tilde{Y}_{1:T}$ and, for all $i,j$, we have   
	\begin{align}
	\CPDF{\tilde{Y}_{1:T}}{\tilde{y}_{1:T}}{\tilde{\Theta}=\tilde{\theta}_j}&=	\CPDF{\tilde{Y}_{1:T}}{\tilde{y}_{1:T}}{\Theta=\theta_i,\tilde{\Theta}=\tilde{\theta}_j}\nonumber\\
	&=p_{\tilde{\theta}_j}\paren{\tilde{y}_1,\dots,\tilde{y}_T},\nonumber
	\end{align}
 which completes the proof.
	\section{Proof of Lemma \ref{Lem: Density}}\label{App: Density}
	Note that the conditional cumulative distribution function of $\tilde{Y}^l_k$ given the event \newline $\left\{\tilde{Y}_k^{1:l-1}=\tilde{y}^{1:l-1}_k,\tilde{Y}_{1:k-1}=\tilde{y}_{1:k-1},\Theta=\theta_i, \tilde{\Theta}=\tilde{\theta}_j\right\}$ can be written as \eqref{Eq: CDF} where ($a$) follows from the definition of $\tilde{Y}^l_k$: 
	\begin{align}
	\tilde{Y}^l_k&=\INVCCDF{{l,k,\tilde{\theta}_j}}{U^l_k}{\tilde{Y}^{1:l-1}_k,\tilde{Y}_{1:k-1}}\nonumber\\
	U^l_k&=\CCDF{l,k,\theta_i}{Y^l_k}{{Y}_k^{1:l-1},Y_{1:k-1}}\nonumber
	\end{align}
	 and  $(b)$ follows from the fact that $\CCDF{l,k,\tilde{\theta}_j}{\cdot}{\cdot}$  is invertible with respect to its first argument. We next show that $U^l_k$ given $\left\{\tilde{Y}_k^{1:l-1}=\tilde{y}^{1:l-1}_k,\tilde{Y}_{1:k-1}=\tilde{y}_{1:k-1},\Theta=\theta_i, \tilde{\Theta}=\tilde{\theta}_j\right\}$ is a uniformly distributed random variable. To this end, let $z$ denote a real number from the interval $\left[0,1\right]$. Then, we have \eqref{Eq: Unifrom} where $(a)$ follows from the fact that $y^{1:l-1}_k,y_{1:k}$ can be uniquely obtained from $\tilde{y}^{1:l-1}_k,\tilde{y}_{1:k}$ since each nonlinear mapping $\CCDF{l,k,\theta}{\cdot}{\cdot}$ is  invertible with respect to its first argument and $(b)$ follows from the Markov chain $\tilde{\Theta}\longrightarrow \paren{\Theta,Y^{1:l-1}_k,Y_{1:k}}\longrightarrow Y^l_k$.
	 
	  Combining \eqref{Eq: CDF} and \eqref{Eq: Unifrom}, we have 
	\begin{align}
	&\CP{\tilde{Y}^{l}_k\leq \tilde{y}^l_k}{\tilde{y}^{1:l-1}_k,\tilde{y}_{1:k-1},\Theta=\theta_i, \tilde{\Theta}=\tilde{\theta}_j}\nonumber\\
	&\hspace{4cm} =\CCDF{l,k,\tilde{\theta}_j}{\tilde{y}^l_k}{\tilde{y}^{1:l-1}_k,\tilde{y}_{1:k-1}}\nonumber
	\end{align}
	which implies that the conditional p.d.f. of $\tilde{Y}^l_k$ given the event \newline $\left\{\tilde{Y}_k^{1:l-1}=\tilde{y}^{1:l-1}_k,\tilde{Y}_{1:k-1}=\tilde{y}_{1:k-1},\Theta=\theta_i, \tilde{\Theta}=\tilde{\theta}_j\right\}$ is given by $\CPDF{\tilde{\theta}_j}{\tilde{y}^l_k}{\tilde{y}_k^{1:l-1},\tilde{y}_{1:k}}$.
	\begin{figure*}
		\begin{align}\label{Eq: CDF}
	\hspace{-1cm}	\CP{\tilde{Y}^{l}_k\leq \tilde{y}^l_k}{\tilde{y}^{1:l-1}_k,\tilde{y}_{1:k-1},\Theta=\theta_i, \tilde{\Theta}=\tilde{\theta}_j}
		&\stackrel{(a)}{=}\CP{\INVCCDF{{l,k,\tilde{\theta}_j}}{U^l_k}{\tilde{y}^{1:l-1}_k,\tilde{y}_{1:k-1}}\leq \tilde{y}^l_k}{\tilde{y}^{1:l-1}_k,\tilde{y}_{1:k-1},\Theta=\theta_i, \tilde{\Theta}=\tilde{\theta}_j}\nonumber\\
		&\stackrel{(b)}{=}\CP{U^l_k\leq \CCDF{l,k,\tilde{\theta}_j}{\tilde{y}^l_k}{\tilde{y}^{1:l-1}_k,\tilde{y}_{1:k-1}}}{\tilde{y}^{1:l-1}_k,\tilde{y}_{1:k-1},\Theta=\theta_i, \tilde{\Theta}=\tilde{\theta}_j}
		\end{align}
		\hrule
		\begin{align}\label{Eq: Unifrom}
	\hspace{-1cm}	\CP{U^l_k\leq z}{\tilde{y}^{1:l-1}_k,\tilde{y}_{1:k-1},\Theta=\theta_i, \tilde{\Theta}=\tilde{\theta}_j} &\stackrel{(a)}{=} \CP{Y^l_k\leq F^{-1}_{\theta_i}\left(\left.{z}\right|{{y}_k^{1:l-1},y_{1:k-1}}\right)}{y^{1:l-1}_k,y_{1:k-1},\Theta=\theta_i, \tilde{\Theta}=\tilde{\theta}_j}\nonumber\\
		&\stackrel{(b)}{=}\CCDF{\theta_i}{F^{-1}_{\theta_i}\left(\left.{z}\right|{{y}_k^{1:l-1},y_{1:k-1}}\right)}{y_k^{1:l-1},y_{1:k-1}}\nonumber\\
		&=z
		\end{align}
		\hrule
	\end{figure*}
	\section{Proof of Theorem \ref{Theo: Conv}}\label{App: Conv}
	The objective function  in \eqref{Eq: OP} can be written as 
	\begin{align}
	&\frac{1}{T} \sum_{k=1} ^T \ES{d\paren{Y_k,\tilde{Y}_k}}\nonumber\\
	&= \frac{1}{T} \sum_{k=1} ^T \sum_{i,j}\CES{d\paren{Y_k,\tilde{Y}_k}}{\Theta=\theta_i,\tilde{\Theta}=\tilde{\theta}_j}P_{ji}\PRP{\Theta=\theta_i}\nonumber\\
	&=\frac{1}{T} \sum_{k=1} ^T\sum_{i,j}\ES{d\paren{Y_k,\NLM{k}{Y_{1:k}}{\theta_i}{\tilde{\theta}_j}}}P_{ji}p_i\nonumber\\
	&=\frac{1}{T} \sum_{k=1} ^T\sum_{i,j}L_k\paren{\theta_i,\tilde{\theta}_j}P_{ji}p_i,
	\end{align}
	where $\NLM{k}{\cdot}{\theta}{\tilde{\theta}}$ is the vector-valued transformation that generates $\tilde{Y}_k$ using $\paren{Y_{1:k},\Theta,\tilde{\Theta}}$ and $L_k\paren{\theta_i,\tilde{\theta}_j}=\ES{d\paren{Y_k,\NLM{k}{Y_{1:k}}{\theta_i}{\tilde{\theta}_j}}}$. Thus, the objective function is linear in the randomization probabilities. Also, it can be shown that the privacy constraint is convex in the optimization variables \cite{CT06}. Thus, the optimization problem in \eqref{Eq: OP} is  convex.
	\section{Proof of Theorem  \ref{Theo: EP-LB}}\label{App: EP-LB}
 Using Fano's inequality \cite{CT06}, we can lower bound the error probability of any estimator of $\Theta$, based on  $\tilde{Y}_{1:T}$,  as 
	\begin{align}\label{Eq: Fano}
	\PRP{\Theta\neq\hat{\Theta}\paren{\tilde{Y}_{1:T}}}&\geq \frac{\CDSE{\Theta}{\tilde{Y}_{1:T}}-1}{\log\abs{\vec{\Theta}}},\nonumber\\
	&\stackrel{(a)}{=} \frac{\DSE{\Theta}-\MI{\Theta}{\tilde{Y}_{1:T}}-1}{\log\abs{\vec{\Theta}}},
	\end{align}  
	where $\CDSE{\Theta}{\tilde{Y}_{1:T}}$ denotes the conditional entropy of $\Theta$ given $\tilde{Y}_{1:T}$ and $(a)$ follows from the definition of the mutual information.
	In Appendix \ref{App: Dist_Gen}, we show that the following Markov chain holds 
	\begin{align}
	\Theta\longrightarrow\tilde{\Theta}\longrightarrow \tilde{Y}_{1:T}\nonumber.
	\end{align}
	Hence, using the data processing inequality \cite{CT06}, we have 
	\begin{align}\label{Eq: MI-UB}
	\MI{\Theta}{\tilde{Y}_{1:T}} \leq  \MI{\Theta}{\tilde{\Theta}}. 
	\end{align}
	Combining \eqref{Eq: MI-UB} and \eqref{Eq: Fano}, we have 
		\begin{align}\
	\PRP{\Theta\neq\hat{\Theta}\paren{\tilde{Y}_{1:T}}}&\geq 
 \frac{\DSE{\Theta}-\MI{\Theta}{\tilde{\Theta}}-1}{\log\abs{\vec{\Theta}}}\nonumber\\
 &\stackrel{(a)}{\geq} 
 \frac{\DSE{\Theta}-I_0-1}{\log\abs{\vec{\Theta}}}\nonumber
	\end{align}  
	where $(a)$ follows from the fact that the mutual information between the private parameter and the pseudo parameter is upper bounded by the leakage level of private information in \eqref{Eq: OP}. 
	
	\section{Proof of Lemma \ref{Lem: Density}}\label{App: GD}
	Note that the distribution of $Y_k$ given $\left\{Y_{1:k-1}=y_{1:k-1},\Theta=\theta_i\right\}$ is Gaussian with mean $\hat{y}_{k\left|k-1\right.}$ and covariance $\Sigma^{o,i}_{k\left|k-1\right.}$ where
	\begin{align}
\hat{y}_{k\left|k-1\right.}&=\CES{Y_k}{y_{1:k-1}}\nonumber \\
&=C_i\hat{x}_{k\left|k-1\right.},\nonumber
\end{align}
and
	\begin{align}
\Sigma^{o,i}_{k\left|k-1\right.}&=\ES{\paren{Y_k-\hat{y}_{k\left|k-1\right.}}\paren{Y_k-\hat{y}_{k\left|k-1\right.}}^\top}\nonumber\\
&=C_i\Sigma^i_{k\left|k-1\right.}C_i^\top+Q^i_v,\nonumber
\end{align}
where $\hat{x}_{k\left|k-1\right.}$ is the optimal predictor of $X_k$ based on \textcolor{black}{$\left\{Y_{1:k-1}=y_{1:k-1},\Theta=\theta_i\right\}$} and $\Sigma^i_{k\left|k-1\right.}$ is the error covariance matrix associated with $\hat{x}_{k\left|k-1\right.}$. Note that the joint p.d.f. of $Y^{1:l}_k$ given $\left\{Y_{1:k-1}=y_{1:k-1},\Theta=\theta_i\right\}$ is also a Gaussian distribution with mean $\hat{y}^{1:l}_{k\left|k-1\right.}$ and covariance $\left[\Sigma^{o,i}_{k\left|k-1\right.}\right]_l$  where  $\left[\Sigma^{o,i}_{k\left|k-1\right.}\right]_{l}$ is a matrix formed by the elements in the first $l$ rows and the first $l$ columns of $\Sigma^{o,i}_{k\left|k-1\right.}$. Thus, using the conditional distribution formula for Gaussian random variables, the  conditional p.d.f. of $Y^l_k$ given $\left\{Y^{1:l-1}_k=y^{1:l-1}_k,Y_{1:k-1}=y_{1:k-1},\Theta=\theta_i\right\}$ is a Gaussian distribution with mean \textcolor{black}{$\hat{\mu}^{l,i}_{lk}$} and variance \textcolor{black}{$\sigma^{i,l}_{lk}$} where 
		\begin{align}
\hat{\mu}^{l,i}_{k}&=\hat{y}^l_{k\left|k-1\right.}-\Delta^{l,i}_k\left[\Sigma^{o,i}_{k\left|k-1\right.}\right]^{-1}_{l-1}\paren{y^{1:l-1}_k-\hat{y}^{1:l-1}_{k\left|k-1\right.}}\nonumber\\
\sigma^{l,i}_{k}&=\Sigma^{l,o,i}_{k\left|k-1\right.}-\Delta^{l,i}_k\left[\Sigma^{o,i}_{k\left|k-1\right.}\right]^{-1}_{l-1}\paren{\Delta^{l,i}_k}^\top\nonumber.
\end{align}
 $\Sigma^{l,o,i}_{k\left|k-1\right.}$ is the $l$th diagonal entry of $\Sigma^{o,i}_{k\left|k-1\right.}$ and $\Delta^{l,i}_k$ is the vector of the first $l-1$ elements in the $l$th row of $\Sigma^{o,i}_{k\left|k-1\right.}$.
\bibliographystyle{IEEEtran}
\bibliography{MTNS}
\begin{IEEEbiography}{Ehsan Nekouei}
	(S’11, M’14 ) is currently an Assistant Professor with the Department of Electrical Engineering, City University of Hong Kong. From 2014 to 2019, he held postdoctoral positions at the KTH Royal Institute of Technology,
	Stockholm, Sweden, and The University of Melbourne, Australia. His current research interests include privacy in networked control systems and integrated processing of human decision-making data.
\end{IEEEbiography}
\begin{IEEEbiography}{Henrik Sandberg}
	(S’01–M’07) received the
	M.Sc. degree in engineering physics and the
	Ph.D. degree in automatic control from Lund
	University, Lund, Sweden, in 1999 and 2004, respectively.
	He is currently a Professor with the Department of Automatic Control, KTH Royal Institute
	of Technology, Stockholm, Sweden. From 2005
	to 2007, he was a Postdoctoral Scholar with
	the California Institute of Technology, Pasadena,
	CA, USA. In 2013, he was a visiting scholar with
	the Laboratory for Information and Decision Systems (LIDS), MIT, Cambridge, MA, USA. He has also held visiting appointments at the Australian National University and the University of Melbourne, Australia.
	His research interests include security of cyberphysical systems, power
	systems, model reduction, and fundamental limitations in control.
	Dr. Sandberg was a recipient of the Best Student Paper Award from
	the IEEE Conference on Decision and Control in 2004 and an Ingvar
	Carlsson Award from the Swedish Foundation for Strategic Research in
	2007. He has been an Associate Editor for the IFAC Journal Automatica and the
	IEEE TRANSACTIONS ON AUTOMATIC CONTROL.
\end{IEEEbiography}

\begin{IEEEbiography}{Mikael Skoglund}
	(S’93–M’97–SM’04-F’19) received
	the Ph.D. degree in information theory from the
	Chalmers University of Technology, Gothenburg,
	Sweden, in 1997.
	In 1997, he joined the Royal Institute of Technology (KTH), Stockholm, Sweden, where he
	was appointed as the Chair in Communication
	Theory in 2003. At KTH, he heads the Communication Theory Division and he is also an
	Assistant Dean in electrical engineering. He is
	also a founding faculty member of the ACCESS
	Linnaeus Center and the Director of the Center Graduate School. He
	has authored and co-authored more than 130 journal and 300 conference papers, and he holds six patents. His research interests include
	problems in source-channel coding, coding and transmission for wireless communications, communication and control, Shannon theory, and
	statistical signal processing.
	Dr. Skoglund has served on numerous technical program committees
	for IEEE sponsored conferences. From 2003 to 2008, he was an Associate Editor for the IEEE TRANSACTIONS ON COMMUNICATIONS, and from
	2008 to 2012, he was on the editorial board of the IEEE TRANSACTIONS
	ON INFORMATION THEORY.
\end{IEEEbiography}
\begin{IEEEbiography}{Karl Henrik Johansson}
	(F’13) received the M.Sc. and Ph.D. degrees from Lund University, Lund, Sweden.
	He is the Director of the Stockholm Strategic Research Area ICT The Next Generation and a Professor
	with the School of Electrical Engineering and Computer Science, KTH Royal Institute of Technology.
	He has held visiting positions with the University
	of California, Berkeley, California Institute of Technology, Nanyang Technological University, HKUST
	Institute of Advanced Studies, and Norwegian University of Science and Technology. His research interests include networked
	control systems, cyber-physical systems, and applications in transportation, energy, and automation. He is a member of the IEEE Control Systems Society
	Board of Governors, the IFAC Executive Board, and the European Control Association Council. He has received several best paper awards and other distinctions. He has been awarded Distinguished Professor with the Swedish Research
	Council and Wallenberg Scholar. He is a recipient of the Future Research Leader
	Award from the Swedish Foundation for Strategic Research and the triennial
	Young Author Prize from IFAC. He is a Fellow of the Royal Swedish Academy
	of Engineering Sciences. He is an IEEE Distinguished Lecturer.
\end{IEEEbiography}

\end{document}